\documentclass[twocolumn,showpacs,aps,prd,superscriptaddress,preprintnumbers,amsmath,amssymb,floatfix]{revtex4}
\usepackage{graphicx}
\usepackage{dcolumn}
\usepackage{bm}

\newcommand{\BABARPubYear}    {06}
\newcommand{\BABARPubNumber}  {025}
\newcommand{\SLACPubNumber} {11816}

\usepackage{relsize}
\usepackage{xspace}
\def\babar{\mbox{\slshape B\kern-0.1em{\smaller A}\kern-0.1em
    B\kern-0.1em{\smaller A\kern-0.2em R}}}
\def\pep2{PEP-II}
\def\D0bar{\kern 0.2em\overline{\kern -0.2em D}{\kern 0.1em}\xspace^0}

\newcommand{\kev}{\ensuremath{\mathrm{\,ke\kern -0.1em V}}\xspace}
\newcommand{\gev}{\ensuremath{\mathrm{\,Ge\kern -0.1em V}}\xspace}
\newcommand{\mev}{\ensuremath{\mathrm{\,Me\kern -0.1em V}}\xspace}
\newcommand{\gevc}{\ensuremath{{\mathrm{\,Ge\kern -0.1em V\!/}c}}\xspace}
\newcommand{\mevc}{\ensuremath{{\mathrm{\,Me\kern -0.1em V\!/}c}}\xspace}
\newcommand{\gevcc}{\ensuremath{{\mathrm{\,Ge\kern -0.1em V\!/}c^2}}\xspace}
\newcommand{\mevcc}{\ensuremath{{\mathrm{\,Me\kern -0.1em V\!/}c^2}}\xspace}
\newcommand{\kevcc}{\ensuremath{{\mathrm{\,ke\kern -0.1em V\!/}c^2}}\xspace}

\def\mum  {\ensuremath{{\,\mu\rm m}}\xspace}
\newcommand{\dedx}{\ensuremath{\mathrm{d}\hspace{-0.1em}E/\mathrm{d}x}\xspace}

\def\piz   {\ensuremath{\pi^0}\xspace}
\def\pip   {\ensuremath{\pi^+}\xspace}
\def\pim   {\ensuremath{\pi^-}\xspace}
\def\pipm  {\ensuremath{\pi^\pm}\xspace}
\def\Kp    {\ensuremath{K^+}\xspace}
\def\Km    {\ensuremath{K^-}\xspace}

\def\Ds    {\ensuremath{D^+_s}\xspace}
\def\Kbar{\kern 0.2em\overline{\kern -0.2em K}{}\xspace}

\def\DsTT{D_{sJ}^*(2317)^+}
\def\DsTO{D_s^{*}(2112)^+}
\def\DsFE{D_{sJ}(2460)^+}
\def\DsTS{D_{s1}(2536)^+}
\def\DsST{D_{s2}(2573)^+}

\def\DsTTdc{D_{sJ}^*(2317)^{++}}
\def\DsTTz{D_{sJ}^*(2317)^0}

\begin{document}

\preprint{BABAR-PUB-\BABARPubYear/\BABARPubNumber}
\preprint{SLAC-PUB-\SLACPubNumber}

\title{\boldmath A Study of the $\DsTT$ and $\DsFE$ Mesons in\\
Inclusive $c\overline{c}$ Production
near $\sqrt{s} = 10.6$~\gev}

%
\author{B.~Aubert}
\author{R.~Barate}
\author{M.~Bona}
\author{D.~Boutigny}
\author{F.~Couderc}
\author{Y.~Karyotakis}
\author{J.~P.~Lees}
\author{V.~Poireau}
\author{V.~Tisserand}
\author{A.~Zghiche}
\affiliation{Laboratoire de Physique des Particules, F-74941 Annecy-le-Vieux, France }
% ...
\collaboration{The \babar\ Collaboration}
\noaffiliation

\date{\today}

\begin{abstract}
A study of the $\DsTT$ and $\DsFE$ mesons in inclusive
$c\bar{c}$ production is presented
using 232~${\rm fb}^{-1}$ of data collected
by the \babar\   experiment near $\sqrt{s} = 10.6$~\gev. 
Final states consisting of a $\Ds$ meson along with one or more 
$\piz$, $\pi^\pm$, or
$\gamma$ particles are considered. Estimates of the
mass and limits on the width are provided for both mesons and
for the $\DsTS$ meson. A search is also performed for neutral and
doubly-charged partners of the $\DsTT$ meson.
\end{abstract}

\maketitle

\section{Introduction}

The $\DsTT$ meson, 
discovered by this collaboration~\cite{Aubert:2003fg}
and confirmed by others~\cite{Besson:2003cp,Abe:2003jk},
does not conform to conventional models of $c\bar{s}$ meson 
spectroscopy. Included with this discovery were suggestions of a 
second state, the $\DsFE$ meson. This meson,
observed by CLEO~\cite{Besson:2003cp} and 
confirmed by this collaboration~\cite{Aubert:2003pe}
and BELLE~\cite{Abe:2003jk}, has a mass that is also lower than
expectations~\cite{Godfrey:1985xj,Godfrey:1991wj,Isgur:1991wq,DiPierro:2001uu}. 
Because the masses of these two states are so unusual
there has been speculation~\cite{Barnes:2003dj,Lipkin:2003zk} 
that both possess an exotic, four-quark component.
The possibility that the $\DsTT$ and $\DsFE$ are exotic
has attracted considerable experimental and theoretical interest
and has focused renewed attention on the subject of charmed-meson
spectroscopy in general.

Presented in this paper is an updated analysis of these two
states using $232$~${\rm
fb}^{-1}$ of $e^+e^- \to c \overline{c}$ data collected by the \babar\ 
experiment. From this analysis  new estimates of the
$\DsTT$ and $\DsFE$ masses, limits on their intrinsic widths,
calculations on their production cross sections,
and the branching ratios of $\DsFE$
decays to $\Ds\gamma$ and $\Ds\pip\pim$ with respect to
its decay to $\Ds\piz\gamma$ are presented. 

These measurements are performed by fitting the invariant mass spectra
of combinations~\cite{cpfootnote} of $\Ds\piz$, 
$\Ds\gamma$, $\Ds\piz\gamma$, $\Ds\piz\piz$, $\Ds\gamma\gamma$, and
$\Ds\pip\pim$ particles. Combinations of $\Ds\pip$ and $\Ds\pim$ are also
studied to search for new states. The mass spectrum of each
final-state combination is studied in detail. In particular,
features in the spectra that arise from
reflections of other $c\bar{s}$ meson decays are individually identified
and modeled. The analysis of the $\Ds\piz\gamma$ final state includes
an explicit search for the two most likely sub-resonant decay channels for
$\DsFE$ meson decay.

This paper is organized as follows. First,
the current status of the $\DsTT$ and $\DsFE$ mesons is reviewed.
The reconstruction of $\Ds$, $\piz$, $\gamma$, and $\pipm$ candidates
is then described, including an estimate of $\Ds$ yield in terms
of the $\Ds\to\phi\pip$ branching fraction. Each combination
of final-state particle species 
is then discussed individually. The paper finishes
with a summary of results and conclusions.

\section{\boldmath Review of the $\DsTT$ and $\DsFE$}

\begin{figure}
\includegraphics[width=\linewidth]{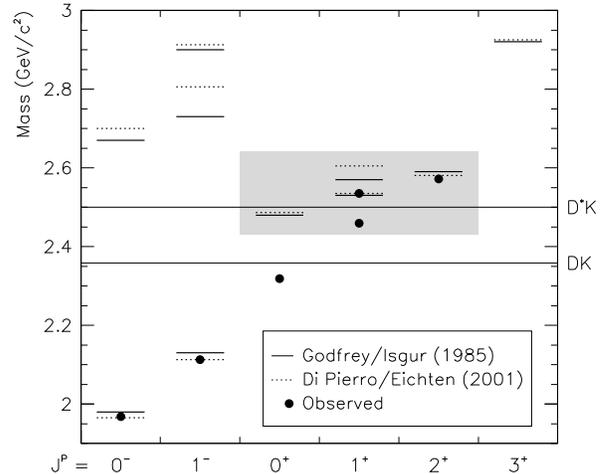}
\vskip -0.25in
\caption{\label{fg:godfreyisgur}The $c\bar{s}$ meson spectrum,
as predicted by Godfrey and Isgur~\cite{Godfrey:1985xj} (solid lines)
and Di Pierro and Eichten~\cite{DiPierro:2001uu} (dashed lines)
and as observed by experiment (points). The $DK$ and $D^*K$ mass
thresholds are indicated by the horizontal lines spanning the width
of the plot.
}
\end{figure}

Much of the theoretical work on the $c\bar{s}$ system 
has been performed in the limit of heavy $c$ quark mass using potential 
models~\cite{Godfrey:1985xj,Godfrey:1991wj,Isgur:1991wq,DiPierro:2001uu}
that treat the 
$c\bar{s}$ system much like a hydrogen atom. 
Prior to the discovery of the $\DsTT$ meson,
such models were successful
at explaining the masses of all known $D$ and $D_s$ states and
even predicting, to good accuracy, the masses of many $D$ mesons 
(including the $\DsTS$ and $\DsST$)
before they were observed (see Fig.~\ref{fg:godfreyisgur}). 
Several of the predicted $D_s$ states were not confirmed experimentally, 
notably the lowest mass $J^P=0^+$ state (at around 2.48~\gevcc)
and the second lowest mass $J^P=1^+$ state (at around 2.58~\gevcc).
Since the predicted widths
of these two states were large, they would be hard to observe,
and thus the lack of experimental evidence
was not a concern.

The $\DsTT$ meson has been observed in the decay 
$\DsTT\to\Ds\piz$~\cite{Aubert:2003fg,Besson:2003cp,Abe:2003jk,Krokovny:2003zq,Aubert:2004pw}.
The mass is measured to be around $2.32$~\gevcc, which is below
the $DK$ threshold. Thus, this particle is forced to decay either
electromagnetically, of which there is no experimental evidence, 
or through the observed isospin-violating $\Ds\piz$ strong decay.
The intrinsic width is small enough that only upper limits have been
measured (the best limit previous to this paper being
$\Gamma<4.6$~\mev\  at 95\% CL as established by BELLE~\cite{Abe:2003jk}).
If the $\DsTT$ is the missing $0^+$ $c\bar{s}$ meson state,
the narrow width could be explained by the lack of an isospin-conserving
strong decay channel. The low mass (160~\mevcc\  below expectations) is more
surprising and has led to the speculation that the $\DsTT$ does not
belong to the $\Ds$ meson family at all but is instead some type of exotic 
particle, such as a four-quark state~\cite{Barnes:2003dj}.

The $\DsFE$ meson has been observed decaying to
$\Ds\piz\gamma$~\cite{Besson:2003cp,Aubert:2003pe,Abe:2003jk,Krokovny:2003zq,Aubert:2004pw},
$\Ds\pip\pim$~\cite{Abe:2003jk}, and
$\Ds\gamma$~\cite{Abe:2003jk,Krokovny:2003zq,Aubert:2004pw}. 
The intrinsic width is small enough that only upper limits have been
measured (the best limit previous to this paper being
$\Gamma<5.5$~\mev\  at 95\% CL as established by BELLE~\cite{Abe:2003jk}).
The $\Ds\gamma$ decay implies a spin of at least one, and so it is natural
to assume that the $\DsFE$ is the missing $1^+$ $c\bar{s}$ meson state.
Like the $\DsTT$, the $\DsFE$ is substantially lower in mass than
predicted for the normal $c\bar{s}$ meson. This suggests that a similar 
mechanism is deflating the masses
of both mesons, or that both the states belong to the same family
of exotic particles.

The spin-parity of the $\DsTT$ and $\DsFE$ mesons has not
been firmly established. The decay mode of the $\DsTT$ alone implies
a spin-parity assignment from the natural $J^P$ series
$\{0^+,1^-,2^+,\dots\}$, assuming parity conservation. Because of the
low mass, the assignment $J^P=0^+$ seems most reasonable, although
experimental data have not ruled out higher spin. 
It is not clear whether electromagnetic decays such as $\DsTO\gamma$ can
compete with the strong decay to $\Ds\piz$, even with isospin violation.
Thus, the absence of experimental evidence for radiative decays such 
as $\DsTT\to\DsTO\gamma$ is not conclusive.

Experimental evidence for the spin-parity of the $\DsFE$ meson
is somewhat stronger. The observation of the decay to $\Ds\gamma$
alone rules out $J=0$. Decay distribution studies in
$B\to\DsFE D_s^{(*)-}$~\cite{Krokovny:2003zq,Aubert:2004pw} favor the 
assignment $J=1$. Decays to either $\Ds\piz$, $D^0 K^+$,
or $D^+K^0$ would be favored
if they were allowed. Since these decay channels are not observed, 
this suggests,
when combined with the other observations, the assignment $J^P = 1^+$.
In this case, the decay to $\DsTT\gamma$ is allowed, but it may be small in
comparison to the $\Ds\gamma$ decay mode.

Table~\ref{tb:modes} lists various possible decay channels
for the $\DsTT$ and $\DsFE$ mesons. Several of these decays are
forbidden assuming the spin-parity assignments discussed above.

\begin{table}
\caption{\label{tb:modes}A list of various decay channels and
whether they have been seen, are allowed,
or are forbidden in the decay of
the $\DsTT$ and $\DsFE$ mesons. The predictions assume 
a spin-parity assignment of $J^P=0^+$ and $1^+$, respectively.}
\begin{ruledtabular}
\renewcommand{\baselinestretch}{1.3}
\begin{tabular}{l@{}lll}
\multicolumn{2}{l}{Decay Channel} & $\DsTT$  & $\DsFE$  \\
\hline
\multicolumn{2}{l}{$\Ds\piz$}             & Seen            & Forbidden \\
\multicolumn{2}{l}{$\Ds\gamma$}           & Forbidden       & Seen      \\
\multicolumn{2}{l}{$\Ds\piz\gamma$ (a)}   & Allowed         & Allowed   \\
             & $\DsTO\piz$                & Forbidden       & Seen      \\
             & $\DsTT\gamma$              & ---             & Allowed   \\
\multicolumn{2}{l}{$\Ds\piz\piz$}         & Forbidden       & Allowed   \\
\multicolumn{2}{l}{$\Ds\gamma\gamma$ (a)} & Allowed         & Allowed   \\
             & $\DsTO\gamma$              & Allowed         & Allowed   \\
\multicolumn{2}{l}{$\Ds\pip\pim$}         & Forbidden       & Seen      \\
\end{tabular}
\end{ruledtabular}
{\footnotesize (a) Non-resonant only\hfill}
\end{table}

\section{The \babar\  Detector and Dataset}
\label{sec:BaBarDetector}

The data used in this analysis were recorded with the \babar\  
detector at the PEP-II asymmetric-energy storage rings
and correspond to an integrated luminosity
of 232~${\rm fb}^{-1}$ collected on or just
below the $\Upsilon(4S)$ resonance.

A detailed description of the \babar\ detector is presented 
elsewhere~\cite{Aubert:2001tu}. 
Charged particles are detected with a five-layer, double-sided
silicon vertex tracker (SVT) and a 40-layer drift chamber (DCH) using a
helium-isobutane gas mixture, placed in a 1.5-T solenoidal field produced
by a superconducting magnet. The  charged-particle momentum resolution 
is approximately $(\delta p_T/p_T)^2 = (0.0013 \, p_T)^2 +
(0.0045)^2$, where $p_T$ is the transverse momentum   in \gevc. 
The SVT, with a typical
single-hit resolution of 10\mum,  measures the impact
parameters of charged-particle tracks in both the plane transverse to
the beam direction and along the beam.
Charged-particle types are identified from the ionization energy loss
(\dedx) measured in the DCH and SVT, and from the Cherenkov radiation
detected in a ring-imaging Cherenkov device (DIRC). Photons are
detected by a CsI(Tl) electromagnetic calorimeter (EMC) with an
energy resolution $\sigma(E)/E = 0.023\cdot(E/\gev)^{-1/4}\oplus
0.019$. 
The return yoke of the superconducting coil is instrumented with
resistive plate chambers (IFR) for the identification of muons and the
detection of neutral hadrons.

\section{Candidate Reconstruction}

The goal of this analysis is to study the possible decay of the
$\DsTT$ and $\DsFE$ mesons into the final states listed 
in Table~\ref{tb:modes}. These decay channels consist of one $\Ds$
meson combined with up to two additional particles selected from
$\piz$, $\pipm$, and $\gamma$. The first step in this analysis is to 
identify $\Ds$ mesons in the \babar\  
data. For each resulting $\Ds$ candidate a search is performed for
associated $\piz$, $\gamma$, and $\pipm$ particles. Signals from
$\DsTT$ and $\DsFE$ decay are isolated using the invariant mass
of the desired combination of particle species.

The $\Ds\to\Kp\Km\pip$ decay mode is used to select a high-statistics
sample of $\Ds$ meson candidates. Each $K^{\pm}$ and $\pi^\pm$ candidate is 
separated from other charged particle species by a likelihood-based
particle identification algorithm based on the Cherenkov-photon
information from the DIRC together with \dedx  measurements from
the SVT and DCH. A geometrical fit to a common vertex
is applied to each $\Kp\Km\pip$ combination.
An acceptable $\Kp\Km\pip$ candidate must have a fit
probability greater than 0.1\% and a trajectory consistent with
originating from the $e^+e^-$ luminous region. 
To reduce combinatorial background, each $\Kp\Km\pip$ candidate
must have a momentum $p^*$ in the $e^+e^-$ center-of-mass frame
greater than 2.2~\gevcc, a requirement that also removes nearly all
contributions from $B$-meson decay. Background from
$D^0\to\Kp\Km$, which is evident from the corresponding $\Kp\Km$
mass distribution, is removed by requiring that the $\Kp\Km$
mass be less than 1.84~\gevcc.

\begin{figure}
\includegraphics[width=\linewidth]{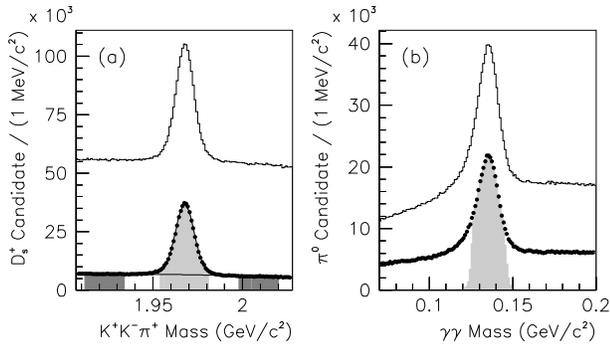}
\vskip -0.10in
\caption{\label{fg:spectrum.samplepaper}(a) The invariant mass
spectrum of $\Ds$ candidates before (top histogram) and after
(bottom points) applying subresonant $\phi\pip$ and $\Kbar^*\Kp$ selection.
The light (dark) areas indicate the signal (sideband) regions.
(b) The invariant $\gamma\gamma$ mass of $\piz$ candidates
before (top histogram) and after (bottom points) applying the 
$\piz$ veto described in the text. The light histogram indicates
those candidates that pass the $\chi^2$ requirement. The curve
in  (a) is the $\chi^2$ fit described in the text.
}
\end{figure}

\begin{figure}
\includegraphics[width=0.8\linewidth]{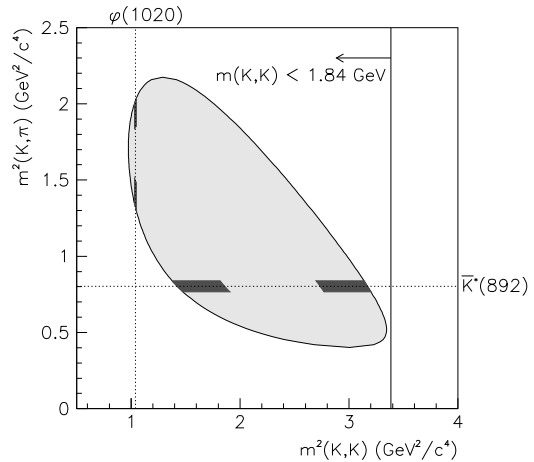}
\vskip -0.20in
\caption{\label{fg:spectrum.dalitz}An illustration of the
$\Ds$ sub-resonant selection requirements.
The light shaded area is the kinematic range for 
$\Ds\to\Kp\Km\pip$ decay.
The vertical line with an arrow represents the selection
requirement used to remove
$D^0\to\Kp\Km$ decay. The four dark regions indicate those portions of the
$\Ds$ phase space used for final candidate selection, 
corresponding to $\Ds\to\phi\pip$ and $\Ds\to \Kbar^*\Kp$ decay.
}
\end{figure}

The upper histogram in Fig.~\ref{fg:spectrum.samplepaper}(a) 
shows the $\Kp\Km\pip$ mass 
distribution for all candidates. A clear $\Ds$ signal is seen. 
To reduce the background further,
only those candidates with $\Kp\Km$ mass
within 10~\mevcc of the $\phi(1020)$ mass
or with $\Km\pip$ mass within 50~\mevcc of the $\Kbar^*(892)$
mass are retained; these densely populated
regions in the $D_s^+$ Dalitz plot do not overlap 
(see Fig.~\ref{fg:spectrum.dalitz}).
The decay products of the vector particles $\phi(1020)$ and 
$\Kbar^*(892)$ exhibit the
expected $\cos^2 \theta_h$ behavior required
by conservation of angular momentum, where $\theta_h$ is the
helicity angle. The signal-to-background ratio
is further improved by requiring $|\cos\theta_h|>0.5$. The lower histogram
of Fig.~\ref{fg:spectrum.samplepaper}(a) 
shows the net effect of these additional selection
criteria. The $D_s^+$ signal ($1.954<m(\Kp\Km\pip)<1.981$~\gevcc)
and sideband ($1.912<m(\Kp\Km\pip)<1.934~\gevcc$ and
$1.998<m(\Kp\Km\pip)<2.020$~\gevcc) regions are shaded.
This distribution can be reasonably modeled in a $\chi^2$ fit
by the sum of two Gaussian distributions with a common mean 
(hereafter referred to as a double Gaussian)
on top of a quadratic background.
The result of this fit is a $\Ds$ signal peak
consisting of approximately 410~000 decays and
a mass of $1967.8$~\mevcc with negligible statistical error.

\begin{figure}
\includegraphics[width=0.8\linewidth]{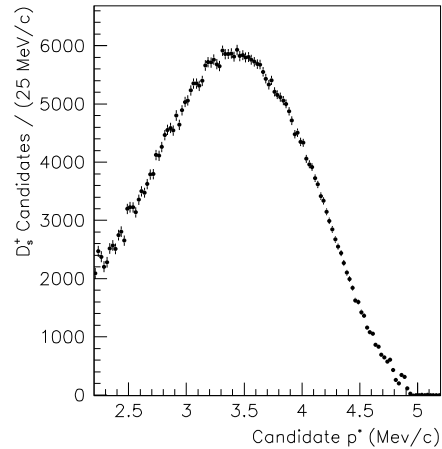}
\vskip -0.20in
\caption{\label{fg:spectrum.spectrum}The sideband subtracted
$p^*$ spectrum for $\Ds$ candidates after all selection requirements
are met.
}
\end{figure}

The approximate $p^*$ distribution for selected
$\Ds$ mesons can be obtained by
simple sideband subtraction, assuming linear background behavior under
the $\Ds$. The result is shown in Fig.~\ref{fg:spectrum.spectrum}.

The final list of $\Ds$ candidates are those that lie within the
signal window. For each such candidate, the momentum vector is
calculated from the simple addition of $\Kp$, $\Km$, and $\pip$
momentum vectors. The energy is chosen to reproduce the
PDG value for the $\Ds$ mass 
($1968.5 \pm 0.5)$~\mevcc~\cite{Eidelman:2004wy}.

\begin{figure}
\includegraphics[width=\linewidth]{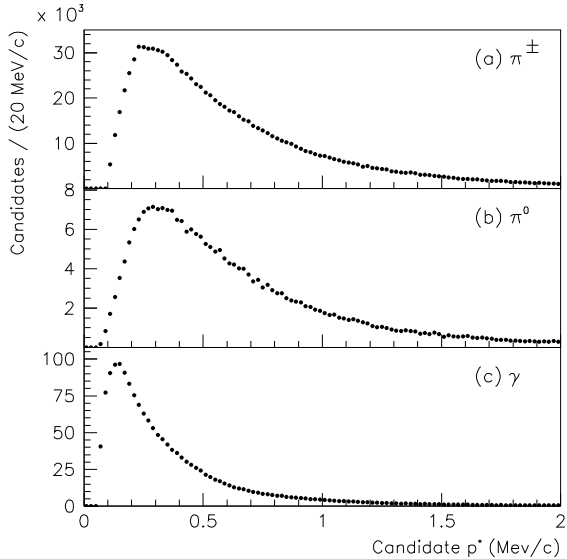}
\vskip -0.15in
\caption{\label{fg:spectrum.spectrum2}The $\Ds$-sideband subtracted
$p^*$ spectrum for associated (a) $\pipm$, (b) $\piz$, and (c) $\gamma$,
after the selection requirements described in the text are fulfilled.
}
\end{figure}

It is assumed that the $\DsTT$ and $\DsFE$ mesons have lifetimes
that are too small to be resolved by the detector. Thus, the most likely
point of $\DsTT$ and $\DsFE$ decay for each $\Ds$ candidate
is chosen to be the interaction point (IP), calculated from the intersection
of the trajectory of the  candidate and the $e^+e^-$ luminous
region. To produce a list of $\pipm$ particles that could arise from
$\DsTT$ or $\DsFE$ decay, the trajectories of
all $\pipm$ candidates that are not daughters of
the $\Ds$ candidate are constrained to the IP using a geometric vertex
fit. The approximate $p^*$ spectrum of these candidates 
associated with real $\Ds$ mesons can be obtained by using simple
$\Ds$ sideband subtraction. The result is shown in 
Fig.~\ref{fg:spectrum.spectrum2}(a).

The selection of $\gamma$ and $\piz$ candidates is a two-step process.
The first step is the selection of a fiducial list of 
$\gamma$ and $\piz$ candidates. The fiducial list of $\gamma$
candidates is constructed from energy clusters in the EMC with
energies above 100~\mev\  and not associated with a charged track.
The energy centroid in the EMC combined with the IP position is used
to calculate the $\gamma$ momentum direction. Each fiducial $\piz$
candidate is constructed from a pair of $\gamma$ particles in the
$\gamma$ fiducial list. This $\gamma$ pair is combined using
a kinematic fit assuming a $\piz$ mass. The resulting $\piz$ momentum
is required to be greater than 150~\mevc. The $\gamma\gamma$ invariant
mass spectrum from this $\piz$ selection is shown in the top histogram
of Fig.~\ref{fg:spectrum.samplepaper}(b). To produce the final fiducial
list of $\piz$ candidates, the $\chi^2$ probability of the kinematic fit is
required to be greater than 2\%.

The final list of $\gamma$ candidates consists of any $\gamma$
in the fiducial list that is not used in the construction of any
$\piz$ in the $\piz$ fiducial list. The final list of $\piz$
candidates consists of any $\piz$ in the fiducial list that does not share
a $\gamma$ with any other $\piz$ in the fiducial list. 
The $\gamma\gamma$ invariant mass distribution of the final list
of $\piz$ candidates is shown in the bottom histograms of
Fig.~\ref{fg:spectrum.samplepaper}(b), both before and after applying
the $\chi^2$ probability requirement. The
approximate $p^*$ spectrum of the final list of $\gamma$ and $\piz$
candidates as determined using $\Ds$ sideband subtraction is
shown in Figs.~\ref{fg:spectrum.spectrum2}(b) and
\ref{fg:spectrum.spectrum2}(c).

\section{Monte Carlo Simulation}

Monte Carlo (MC) simulation is used for the following purposes in this paper:
\begin{itemize}
\item
To calculate signal efficiencies.
\item
To provide independent estimates of background levels.
\item
To characterize the reconstructed mass distribution of the signal.
\item
To predict the behavior of various specific types of backgrounds
(commonly referred to as {\em reflections})
produced when the mass distribution from an established decay mode
is distorted by the loss of one or more final-state particles or
the addition of one or more unassociated particles.
\end{itemize}
Various sets of MC events were generated.
For the purposes of understanding signal efficiencies, signal shapes,
and reflections, individual MC sets of 500~000 decays were generated
for each known decay mode of the $\DsTT$ and $\DsFE$ mesons. In addition,
MC sets of 250~000 decays were produced for each hypothetical 
$\DsTT$ and $\DsFE$ decay
as needed. Finally, a set of $e^+e^-\to c\bar{c}$ events,
corresponding to an integrated luminosity of approximately 80~${\rm fb}^{-1}$,
was generated to study sources of combinatorial background.

Each MC set was processed by the same reconstruction and selection
algorithms used for the data. Independent tests of 
the detector simulation have demonstrated an accurate reproduction
of charged particle detection efficiency. The systematic uncertainty
from these tests is estimated to be 1.3\% for each charged track.
Since the decay $\Ds\to\Kp\Km\pip$ involves three charged tracks,
the systematic uncertainty in $\Ds$ efficiency from the
simulation alone is estimated to be 3.9\%.

The simulation of $\Ds\to\Kp\Km\pip$ decay was designed to
match approximately the known Dalitz structure. MC events were reweighted
to match more precisely the relative  
$\phi\pip$ and $\Kbar^*\Kp$ yields observed in the data. 

The simulation assumes a $\DsTT$ and $\DsFE$
intrinsic width of $\Gamma = 0.1$~\mev. All
$\DsTT$ and $\DsFE$ final states are generated using phase space.
The generated $p^*$ distribution of $\DsTT$ and $\DsFE$ mesons 
in the MC simulation was 
adjusted to roughly reproduce observations.

\section{\boldmath Absolute $\Ds$ Yield}
\label{sec:dsyield}

In order to calculate $\DsTT$ and $\DsFE$ production cross sections,
it is necessary to provide an estimate of absolute $\Ds$ selection
efficiency.
Since this analysis uses $\Ds\to\Kp\Km\pip$ decay, the approach
is to normalize $\Ds$
yield with respect to the $\Ds\to\phi\pip, \phi\to\Kp\Km$
branching fraction, the world average of which is
$1.8\pm 0.4$\%~\cite{Eidelman:2004wy}. To perform this normalization
correctly, the following must be accounted for:
\begin{itemize}
\item
The $\Kbar^*\Kp$ portion of the $\Ds$ sample.
\item
Non-resonant $\Kp\Km\pip$ background under the $\phi$ peak.
\item
The fraction of the $\phi$ signal that falls outside of the $\Kp\Km$ mass 
selection and $\theta_h$ requirements.
\end{itemize}

The $\Kbar^*\Kp$ selection represents approximately 48\% of the 
total $\Ds$ sample. An inspection of the $p^*$ distribution of the
$\phi\pip$ and $\Kbar^*\Kp$ subsamples indicates that this fraction
is, to a good approximation, independent of $p^*$. Therefore, a
constant factor is sufficient to account for the 
contribution from the $\Kbar^*\Kp$ portions of the $\Ds$ sample.

\begin{figure}[!htb]
\begin{center}
\includegraphics[width=\linewidth]{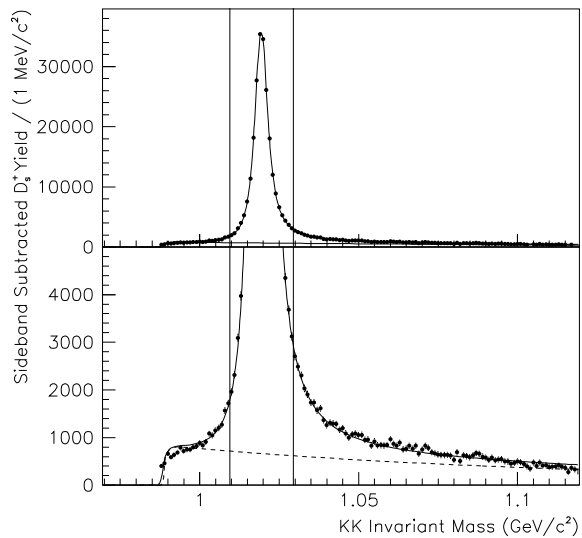}
\vskip -0.15in
\caption{The $\Ds$-sideband subtracted $\Kp\Km$ invariant mass spectrum 
near the $\phi$ mass for the $\Ds$ sample obtained before applying the
$\phi\pip$ and $\Kbar^*\Kp$ selection requirements. The curve is the 
fit described in the text. The dashed line is the portion of the fit
attributed to contributions from other than $\phi$ decay.
The vertical lines indicate the $\phi$ mass selection window.
}
\label{fg:spectrum.fitphipaper}
\end{center}
\end{figure}

Shown in Fig.~\ref{fg:spectrum.fitphipaper} is the $\Ds$-sideband
subtracted $\Kp\Km$ invariant mass spectrum
for all $\Ds$ candidates before applying the $\phi\pip$ and $\Kbar^*\Kp$
selection requirements. A prominent $\phi$ peak is observed.
A binned $\chi^2$ fit to this spectrum is used to extract both the 
fraction of $\phi\pip$ decays that fall outside the
$\phi$ selection window and the number of non-$\phi$ decays that
leak inside. This fit is described below.

To model the signal portion of the $\Kp\Km$ mass spectrum,
the $\phi\to \Kp\Km$ line shape $\sigma(m)$ can be
reasonably well described 
(ignoring potential interference effects)
by a relativistic Breit-Wigner function:
\begin{equation}
\sigma(m) = \frac{ m \Gamma(m) }
                 {\left( m^2 - m^2_0 \right)^2 + m_0^2 \Gamma_\text{tot}^2(m) } \;,
\label{eq:bw}
\end{equation}
where $m_0 = (1019.456\pm0.020)$~\mevcc\  is the intrinsic 
$\phi$ mass~\cite{Eidelman:2004wy}.
The mass-dependent width $\Gamma(m)$ can be approximated by:
\begin{equation}
\Gamma(m) = 0.493 \Gamma_0 \frac{m_0}{m} \left( \frac{q}{q_0} \right)^3 
\frac{1 + q_0^2 R^2}{1 + q^2 R^2}\;,
\end{equation}
where $\Gamma_0 = 4.26$~\mev\  is the intrinsic width,
0.493 is the branching fraction for this decay mode,
$R=3$~$\gev^{-1}$  is an effective $\phi$ radius (to control the tails),
and $q$ ($q_0$) is the total three-momentum of the $K^\pm$ decay products
in the $\phi$ center-of-mass frame assuming an 
effective $\phi$ mass of $m$ ($m_0$):
\begin{equation}
q = \sqrt{ m^2/4 -  m_K^2 } \quad 
\text{and}\quad q_0 = \sqrt{ m^2_0/4 - m_K^2  } \;.
\end{equation}

A reasonable approximation of the total width includes the three
dominant decay modes ($K^+K^-$, $K_S K_L$, $\pip\pim\piz$):
\begin{equation}
\Gamma_\text{tot}(m) = \Gamma(m) + \Gamma'(m) + 0.155\Gamma_0 \;,
\end{equation}
where $\Gamma'(m)$ is calculated in the same manner as $\Gamma(m)$
but using the $K_S/K_L$ mass and branching fraction (0.337) 
and the width for the three-pion decay
mode (which is well above threshold) is treated as a constant.

The fit function $P(m)$ 
used to describe the $\Kp\Km$ mass spectrum 
of Fig.~\ref{fg:spectrum.fitphipaper} is:
\begin{equation}
P(m) = S_\phi(m) + C(m) \;,
\label{eq:phifit}
\end{equation}
where $S_\phi(m)$ is the $\phi$ signal shape and $C(m)$
represents non-resonant contributions. 
The empirical form used for $C(m)$
is a four-parameter threshold function:
\begin{eqnarray}
C(m) &=& \left\{
\begin{array}{ll}
0 & m < a_1 \\
C'(m) & m \ge a_1
\end{array} \right.\label{eq:cthres}\\
C'(m) &=&
\left( 1 + m - a_1 \right)^{a_2}
\left( 1 - \exp( -\left(m - a_0)/a_3 \right) \right) \;. \nonumber
\end{eqnarray}
The form used for $S_\phi(m)$ is the Breit-Wigner 
function of Eq.~\ref{eq:bw}
smeared by a Gaussian:
\begin{equation}
S_\phi(m) = \frac{N}{\delta\sqrt{2\pi}} \int \sigma(m') \:
e^{ -\left(m-m'\right)^2/2\delta^2 }
\:{\rm d}m' \;,
\end{equation}
where $N$ is an overall normalization. In the fit to this function,
the value of $\Gamma_0$ is kept fixed to the PDG value but $m_0$
and $\delta$ are allowed to vary.

Despite its simplicity, the function of Eq.~\ref{eq:phifit} describes
the $\Kp\Km$ spectrum quite well, as shown 
in Fig.~\ref{fg:spectrum.fitphipaper}. The value of $m_0$ produced
by the fit is slightly lower [($-56 \pm 6$)~\kevcc,
statistical error only] than the PDG average~\cite{Eidelman:2004wy}. 
To determine a
correction factor for the $\Ds$ yield, the total $\phi$ yield
(calculated from the integral of the signal line shape determined
by the fit up to a $\Kp\Km$ mass of 1.1~\gevcc)
can be compared to the total number of candidates which fall inside the
$\phi$ mass window. The result is a correction factor of 1.09,
with negligible statistical uncertainty.

To test the above calculation, the fit is repeated on a $\Ds$ sample
that includes the $|\cos\theta_h| > 0.5$ requirement discussed earlier
for the $\phi$. 
The change in $\phi$ integrated yield is consistent with a $\cos^2\theta_h$
distribution. The measured correction factor increases to 1.10.
The difference between this value and $1.09$ 
is taken as a systematic uncertainty.

Other systematic checks performed include 
increasing the range in
$\Kp\Km$ mass of the $\phi$ line shape integration and changing
the value of $R$ from 1 to 5~\gevcc. The total uncertainty in the 1.09
correction factor is found to be a 0.043 (3.9\% relative), as
calculated from a quadrature sum.

\section{Candidate Selection Optimization}

This paper explores the eight final-state combinations
shown in Table~\ref{tb:optimization}, each
involving a $\Ds$ meson and up to two total of
$\pipm$, $\piz$, and/or $\gamma$ particles. 
For each combination it is necessary to distinguish possible signals
from $\DsTT$ and $\DsFE$ decay from combinatorial background.
The separation of signal and background 
is made more distinct if additional candidate selection requirements
are imposed. This section discusses those additional requirements.

In four of the final-state combinations 
($\Ds\piz$, $\Ds\piz\gamma$, $\Ds\gamma$, and $\Ds\pip\pim$) 
a signal is expected.
An estimate of signal significance is calculated in these cases
based on expected signal and background rates, the former
calculated from previously published branching ratio measurements 
combined with the appropriate MC sample. 
For the remainder of the
final states, an estimate of signal sensitivity is calculated
for the hypothetical $\DsTT$ and $\DsFE$ meson decay. This sensitivity
calculation is based on signal efficiency, determined using
MC samples, and expected background levels.

To avoid potential
biases, both the signal significance and sensitivity estimates are 
calculated solely using MC samples.

The following selection requirements are adjusted in order to
produce optimal values of signal significance and sensitivity:
\begin{itemize}
\item
A minimum total center-of-mass momentum $p^*$.
\item
A minimum energy (for $\gamma$) and/or momentum
calculated in the laboratory (for $\piz$) or
center-of-mass (for $\pipm$) frame of reference.
\end{itemize}
For the minimum $p^*$,
it is important to choose the same value for all final-state
combinations in order to minimize systematic uncertainties in the
branching ratios. A minimum value of $p^* > 3.2$~\gevc is chosen as
a reasonable compromise. Values for the remaining selection requirements
are chosen separately for each final-state combination.
The results are listed in Table~\ref{tb:optimization}.

The MC samples can be used to
estimate the approximate efficiency for detecting a signal with a $p^*$
of at least 3.2~\gevc\  after
applying the above selection requirements. The resulting efficiencies
vary between 0.4\% and 13\% (see Table~\ref{tb:optimization}).

\begin{table}
\caption{\label{tb:optimization}Selection requirements for
the final states studied in this paper,
the resulting number of events, and the approximate
efficiency for a $\DsFE$ signal.
The selection requirements are specified 
either in the laboratory (Lab) or center-of-mass (CMS) coordinate
systems.}
\begin{ruledtabular}
\renewcommand{\baselinestretch}{1.3}
\begin{tabular}{lcccrr}
& \multicolumn{3}{c}{Minimum Requirements} & &\\
\cline{2-4}
            & $\gamma$ Energy & $\pi^0$ Mom. & $\pi^\pm$ Mom. & &\\
            & Lab             & Lab          & CMS      & 
\multicolumn{1}{c}{Sample} & Effic.\\
Final State & (\mev)          & (\mevc)      & (\mevc)            & 
\multicolumn{1}{c}{Size} & (\%)\\
\hline
$\Ds\piz$          & --- & 350 & --- &  87 320 &  6.4\\
$\Ds\gamma$        & 500 & --- & --- & 133 398 & 12\\
$\Ds\piz\gamma$    & 135 & 400 & --- & 170 341 &  2.4\\
$\Ds\piz\piz$      & --- & 250 & --- &  17 437 &  0.4\\
$\Ds\gamma\gamma$  & 170 & --- & --- & 575 765 &  7.9\\
$\Ds\pip$          & --- & --- & 300 & 143 149 & 13\\
$\Ds\pim$          & --- & --- & 300 & 219 466 & 13\\
$\Ds\pip\pim$      & --- & --- & 250 & 154 496 &  6.8\\
\end{tabular}
\end{ruledtabular}
\end{table}

\section{\boldmath Cross Section Notation}

To report production yields of a particular
$c\bar{s}$ meson $D_Y$ to a particular final state $\Ds X$, 
the following quantity $\bar{\sigma}(D_Y\to\Ds X)$ is defined:
\begin{eqnarray}
\bar{\sigma}(D_Y\to\Ds X) 
&\equiv& \sigma(e^+e^-\to D_Y ,p^*_Y>3.2~\gevc) \nonumber \\
&\times& \mathcal B( D_Y \to  \Ds X ) \nonumber \\
&\times& \mathcal B( \Ds \to \phi\pip, \phi\to \Kp\Km) \;,
\label{eq:sigmabar}
\end{eqnarray}
where the cross section $\sigma$ is defined for
a center-of-mass momentum $p^*$ above 3.2~\gevc. 
The quantity $\bar{\sigma}$
is calculated by taking the number of $D_Y\to \Ds X$ decays observed in the
data, correcting for efficiency using the appropriate MC sample
(restricted to $p^* > 3.2$~\gevc),
correcting for the relative $\Ds\to\phi\pip$ yield as calculated in
Section~\ref{sec:dsyield}, and dividing by the luminosity
(232~${\rm fb}^{-1}$).
A relative systematic uncertainty of 1.2\% is introduced to account
for the uncertainty in the absolute luminosity.

There is no attempt to correct the cross section
for radiative effects (such as initial-state radiation). Since a
reasonably accurate representation
of such radiative effects is included in our MC samples,
the calculation of selection efficiencies from these samples is
accurate enough for the purposes of this paper.

\section{\boldmath The $\Ds\piz$ Final State}
\label{sec:dspi0}

\begin{figure}
\includegraphics[width=\linewidth]{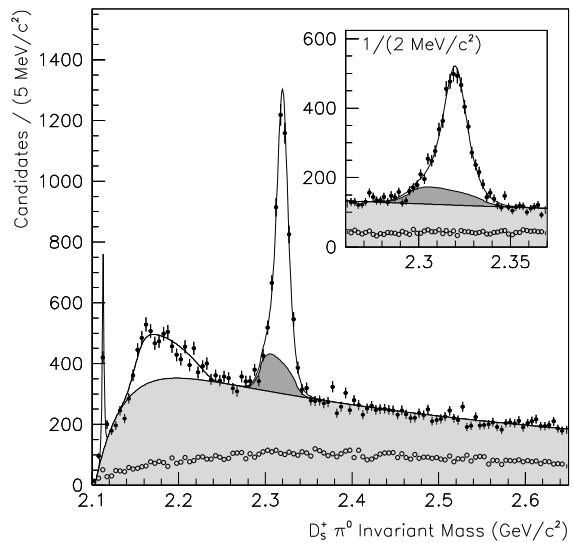}
\vskip -0.15in
\caption{\label{fg:dspi0.paperfitfig2}The invariant mass
distribution for (solid points) $\Ds\piz$ candidates and (open points)
the equivalent using the $\Ds$ sidebands.
The curve represents the likelihood fit described in
the text. Included in this fit is (light shade) a contribution
from combinatorial background and (dark shade) the reflection from
$\DsFE\to\DsTO\piz$ decay. The insert highlights
the details near the $\DsTT$ mass.
}
\end{figure}

Shown in Fig.~\ref{fg:dspi0.paperfitfig2} is the invariant mass
distribution of the $\Ds\piz$ combinations 
after all selection requirements are fulfilled. Signals from $\DsTO$
and $\DsTT$ decay are evident. An unbinned likelihood fit is applied
to this mass distribution in order to extract the parameters and
yield of the $\DsTT$ signal and upper limits on $\DsFE$ decay. The likelihood 
fit includes six distinct sources of $\Ds\piz$ combinations:
\begin{itemize}
\item
$\DsTT\to\Ds\piz$ decay.
\item
$\DsFE\to\Ds\piz$ decay (hypothetical).
\item
$\DsTO\to\Ds\piz$ decay.
\item
A reflection from $\DsTO\to\Ds\gamma$ decay in which an unassociated
$\gamma$ particle is added to form a false $\piz$ candidate.
\item
A reflection from $\DsFE\to\DsTO\piz$ decay in which 
the $\gamma$ from the $\DsTO$ decay is missing.
\item
Combinatorial background from unassociated $\Ds$ and $\piz$ mesons.
\end{itemize}
The probability density function (PDF) used to describe the
mass distribution of each of these sources is described below.

\begin{figure}
\begin{center}
\includegraphics[width=0.49\linewidth]{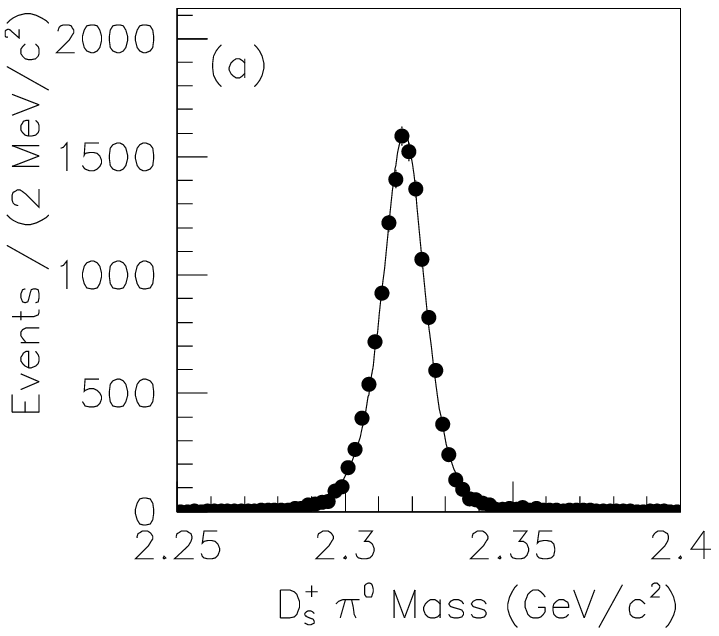}
\includegraphics[width=0.49\linewidth]{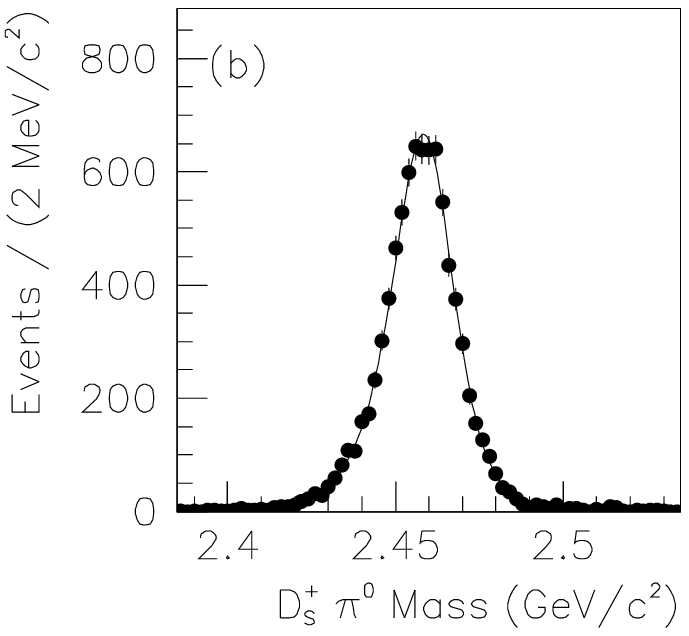}
\includegraphics[width=0.49\linewidth]{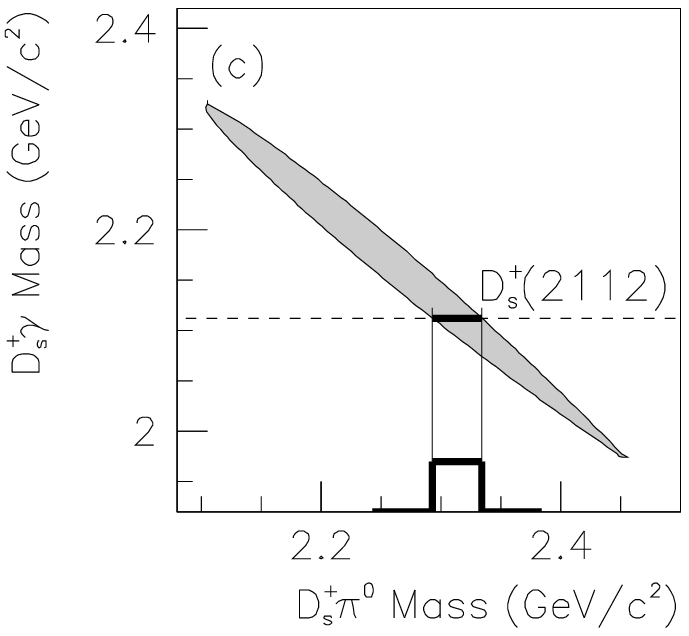}
\includegraphics[width=0.49\linewidth]{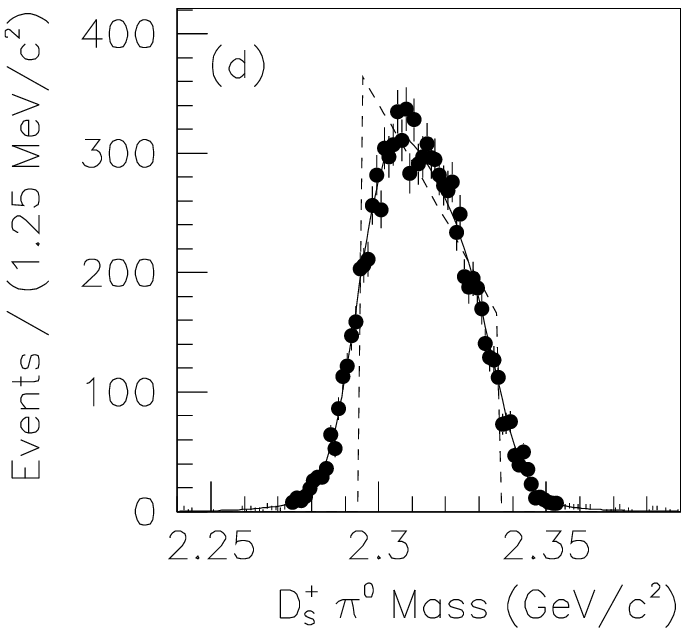}
\end{center}
\vskip -0.25in
\caption{\label{fg:dspi0.signals}The reconstructed $\Ds\piz$ invariant mass
spectrum from (a) $\DsTT$ and (b)
$\DsFE$ MC samples. 
The fit function of Eq.~\ref{eq:modlorentz} is overlaid.
(c) The projection
of $\Ds\piz$ mass for subresonant $\DsFE$ decay through the $\DsTO$ meson
is restricted to a narrow range centered around 2313.4~\mevcc. 
(d) The reconstructed $\Ds\piz$ invariant mass spectrum
for the $\DsFE$ reflection from MC simulation. The solid
curve is the fit function. The dashed curve is the same
fit function with the Gaussian smearing removed.
}
\end{figure}

As shown in Fig.~\ref{fg:dspi0.signals}(a), the reconstructed
mass distribution of the $\DsTT\to\Ds\piz$ decay, as predicted by
MC, has non-Gaussian tails and is slightly asymmetric. To describe this 
shape, the MC sample is fit to a modified Lorentzian function
$F_L(m)$:
\begin{eqnarray}
F_L(m) &=& a_3 \frac{ \left| 1 + a_4 \delta  + a_5 \delta^3 \right| }
                  { \left( 1 + \delta^2 \right)^{a_6} } 
\label{eq:modlorentz}\\
\delta &\equiv& \left( m - a_1 \right)/a_2 \;, \nonumber
\end{eqnarray}                  
where $a_1$ and $a_2$ correspond roughly to a mean and width,
respectively.  This function is simply a convenient parameterization 
of detector resolution.
The fit results are shown in Fig.~\ref{fg:dspi0.signals}(a).
A similar procedure is used for the hypothetical $\DsFE\to\Ds\piz$
decay (Fig.~\ref{fg:dspi0.signals}(b)).

The 350~\mevc\  $\piz$ momentum requirement removes the majority
of $\DsTO\to\Ds\piz$ decays. The remaining signal is modeled using
a distribution $J(m)$ of Gaussians:
\begin{equation}
J(m) = a_1 \int_{a_3}^{a_3a_4}
\frac{1}{a_4 \sigma^2} 
\!\exp\left[ -(m - a_2)^2/2\sigma^2 \right] \text{d}\sigma \;.
\label{eq:gerry}
\end{equation}
The parameters $a_3$ and $a_4$ of this function are 
determined using a fit to a suitable
MC sample. The mean mass $a_2$ is set equal to 
$2112.9$~\mevcc\  (0.5~\mevcc
higher than the PDG value~\cite{Eidelman:2004wy}) to match the data.

A reflection in $\Ds\piz$ produced by $\DsTO\to\Ds\gamma$ decay
appears as a broad distribution peaking at a mass of approximately
2.17~\gevcc. This reflection is produced by fake $\piz$ candidates
consisting of the $\gamma$ particle from $\DsTO$ decay combined with
unassociated $\gamma$ candidates. Kinematics limit this reflection 
to $\Ds\piz$ masses above the quadrature sum
of $\Ds$ and $\piz$ meson masses (approximately 2.1167~\gevcc).
This distribution falls gradually as
mass is increased due to the rapidly falling inclusive 
$\gamma$ energy spectrum. 
Detector resolution tends to smear the lower kinematic mass limit.
To model this reflection, a quadratic function with a sharp lower mass
cut off is convoluted with a Gaussian distribution. The parameters
of this function are
determined directly from the $\Ds\piz$ data sample.

The $\DsFE$ reflection requires careful attention because it appears
directly under the $\DsTT$ signal.
This reflection is produced by the $\Ds\piz$
projection of $\DsFE\to\DsTO\piz$ decay (in which the $\gamma$
from $\DsTO$ decay is 
ignored). A kinematic calculation (Fig.~\ref{fg:dspi0.signals}(c))
of the $\DsFE$ Dalitz distribution predicts that this reflection,
at the limit of perfect resolution and efficiency, 
is a flat distribution in mass squared
centered at a $\Ds\piz$ mass of 2313.4~\mevcc\  with a full width
of 41.3~\mevcc\  (assuming a $\DsFE$ mass of 2458.0~\mevcc).

The $\DsFE$ reflection is flat in mass squared only if the $\piz$
efficiency is constant. In practice this is not the case, as
illustrated by MC simulation (Fig.~\ref{fg:dspi0.signals}(d)). 
To accommodate the non-constant
efficiency, the $\Ds\piz$ mass distribution from the MC sample is fit to a 
function consisting of a bounded quadratic function
smeared by a double Gaussian.
The result of the fit is shown in Fig.~\ref{fg:dspi0.signals}d.

The threshold function $C(m)$ of Eq.~\ref{eq:cthres} is
used to represent the mass spectrum
from combinatorial background where the threshold value 
$a_1$ is fixed to 2103.5~\mevcc, the sum
of the assumed $\Ds$ and $\piz$ masses. The remaining parameters
of $C(m)$ are determined directly from the data.

The results of the likelihood fit to the $\Ds\piz$ mass spectrum
is shown in Fig.~\ref{fg:dspi0.paperfitfig2}. In this fit, the
size, shape, and mean mass of the $\DsFE$ reflection are fixed to
values consistent with the yield and mass results determined 
in Section~\ref{sec:dspi0gam} of this paper. The yield of 
$\DsTO$ and $\DsTT$ decay and the $\DsTT$ mass
is allowed to vary to best match the data. A $\DsTT$ mass
of $(2319.6 \pm 0.2)$~\mevcc\  is obtained (statistical error only).
A total of $3180\pm 80$ $\DsTT$ decays are found.

The fit includes a hypothetical contribution from 
$\DsFE\to\Ds\piz$ in the form of a line shape of fixed shape and
mass. The result is a yield of $-40\pm 50$ (statistical errors only).
The size of this yield is small enough that the curve cannot be
distinguished in Fig.~\ref{fg:dspi0.paperfitfig2}.

\begin{figure}
\begin{center}
\includegraphics[width=\linewidth]{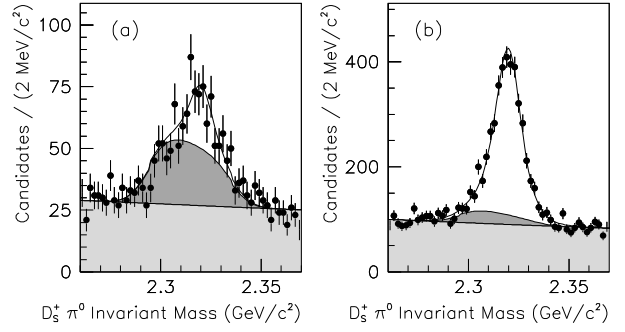}
\end{center}
\vskip -0.25in
\caption{\label{fg:dspi0.plotfitdatadstarpaper}Fit results near the
$\DsTT$ peak for the $\Ds\piz$ sample divided into combinations (a) with
and (b) without a $\Ds$ consistent with $\DsTO\to\Ds\gamma$ decay.
The curves and shaded regions, as described in
Fig.~\ref{fg:dspi0.paperfitfig2}, represent the result of a likelihood fit.
}
\end{figure}

The $\DsFE$ reflection arises from contamination from 
$\DsTO$ decay. It is an interesting exercise to identify and
separate some of this background. This can be accomplished by searching
for any $\gamma$ candidates that, when combined with the $\Ds$
in the same event,
produce a $\Ds\gamma$ mass within 15~\mevcc\  of the $\DsTO$ mass.
Those $\Ds\piz$ combinations in which such a match is not found will
contain a smaller proportion of $\DsFE$ reflection, whereas the remaining
$\Ds\piz$ combinations will contain fewer $\DsTT$ decays. This is indeed
the case as illustrated in Fig.~\ref{fg:dspi0.plotfitdatadstarpaper}.
The same likelihood fit procedure used for the entire sample 
is repeated for these $\Ds\piz$ subsamples, 
including the MC prediction of the yield and shape
of the $\DsFE$ reflection. The fit results are consistent with the data.

As can be seen in Figs.~\ref{fg:dspi0.paperfitfig2} and
\ref{fg:dspi0.plotfitdatadstarpaper}, the $\DsTT$ line shape
derived from MC simulation and used unchanged in the likelihood
describes the data well. Since the MC simulation is
configured with an intrinsic width (0.1~\mev) nearly indistinguishable
from zero, it follows that
the data are consistent with a zero width $\DsTT$ meson.

To extract a 95\% CL upper
limit on the intrinsic $\DsTT$ width, the $\DsTT$ line shape is
convolved with a relativistic
Breit-Wigner function $\sigma_J(m)$ with constant width $\Gamma$:
\begin{equation}
\sigma_J(m) \propto \frac{m_0 \Gamma}{ \left( m_0^2 - m^2 \right)^2
                                + m_0^2 \Gamma^2 } \;.
\label{eq:cbw}
\end{equation}
The fit is then repeated in its entirety at incremental steps
in $\Gamma$ to produce a likelihood curve. Integrating this curve
as a function of $\Gamma$ produces a 95\% CL upper limit of
$\Gamma < 1.9$~\mev (statistical error only).

In order to
produce an estimate of $\DsTT$ yield, the 
fit results must be corrected for selection efficiency.
This efficiency is calculated using
a $\DsTT$ MC sample and is $p^*$ dependent. Since the $p^*$ distribution
observed in data does not exactly match the MC simulation, it is important
to take into account this $p^*$ dependence. Two methods are used to do this.
The first is to weight each $\Ds\piz$ combination by the inverse
of the selection efficiency before applying a likelihood fit.
After correcting for absolute $\Ds\to\phi\pip$ yield
(Section~\ref{sec:dsyield}), the result is:
\begin{equation*}
N(\DsTT\to\Ds\piz,\Ds\to\phi\pip) = 26~290 \pm 650 
\end{equation*}
for $p^* > 3.2$~\gevc\ (statistical error only). 

\begin{figure}
\begin{center}
\includegraphics[width=0.8\linewidth]{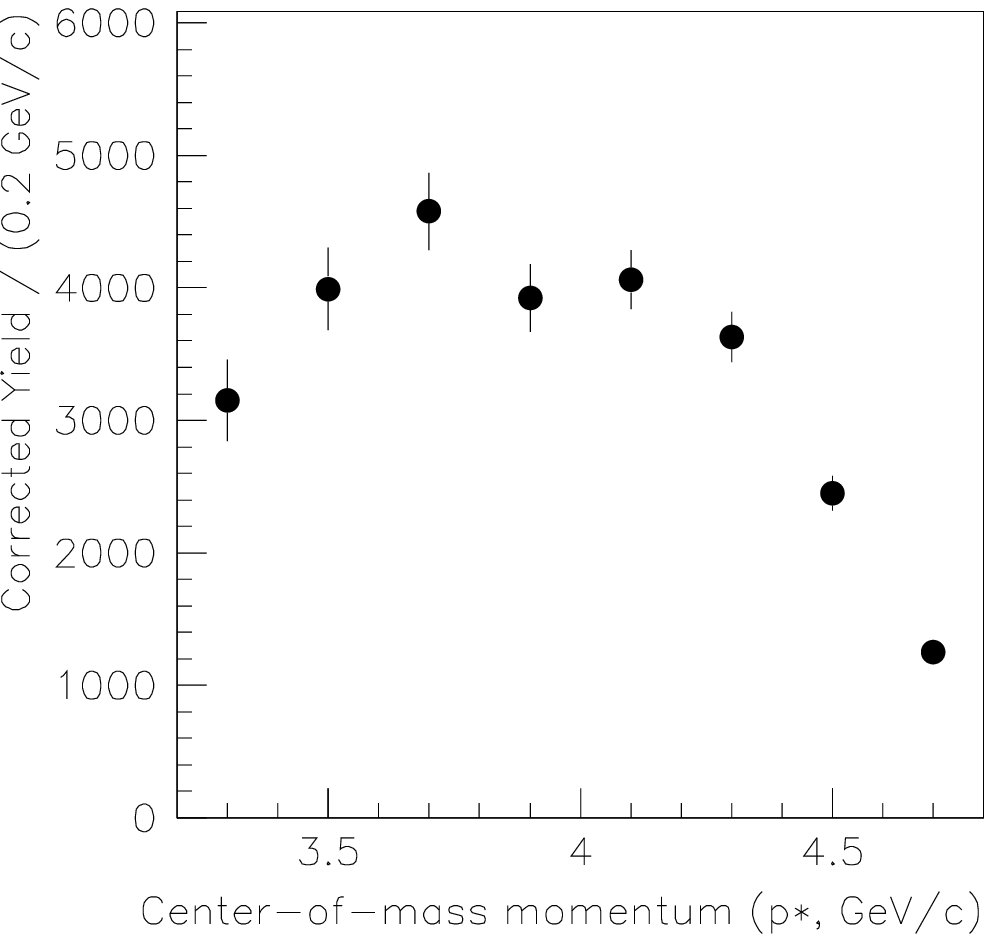}
\end{center}
\vskip -0.25in
\caption{\label{fg:dspi0.pstarpaper}Corrected $\DsTT\to\Ds\piz$ yield
as a function of $p^*$.
}
\end{figure}

The second method
is to divide the $\Ds\piz$ sample into bins of $p^*$. A likelihood fit
is applied to each bin and the yield corrected for the average
selection efficiency in that bin. The result is the $p^*$ distribution
shown in Fig.~\ref{fg:dspi0.pstarpaper}. The total yield from this
method is $26470 \pm 660$ (statistical error only).

The systematic uncertainties for the $\DsTT$ mass and yield
are summarized in Table~\ref{tb:dspi0.syst}. The uncertainties
in $\DsFE$ yield are calculated in the same fashion.
The assumed $\Ds$ mass value and the 1\% relative uncertainty in the
EMC energy scale are the two largest contributors to the error on
the $\DsTT$ mass.
Uncertainties in the signal shape produce the largest uncertainties
in $\DsTT$ yield and width.

For example, the amount of $\DsFE$ reflection is proportional to the
$\DsFE$ yield, which, as will be discussed in Section~\ref{sec:dspi0gam},
has an 11\% uncertainty. Adjusting the contribution 
of this reflection in the fit
with no other changes produces a relative uncertainty of 2.0\% in
the $\DsTT$ yield with little change in the $\DsTT$ mass and width limit.
If the likelihood fit is allowed to choose a $\DsFE$ reflection yield
that best matches the data, little change in either the $\DsTT$ mass
or yield is observed. The limit on the intrinsic width, however,
increases to 3.4~\mev, since reducing the $\DsFE$ reflection allows
the observed mass line shape to accommodate a larger intrinsic $\DsTT$ width.

Another uncertainty that has a similar effect is the assumed
$\Ds\piz$ mass resolution. Small variations in resolution, consistent
with comparisons of data and MC simulation of other known particles,
can change the $\DsTT$ line shape sufficiently to lower or raise
yields by 3.2\%. Allowing better reconstructed resolution provides more
room for a large intrinsic width, raising the 95\% CL for $\Gamma$ to
3.0~\mev.

Other uncertainties in the $\DsTT$ yield include the accuracy 
($\pm 3$\%) of the MC prediction of $\piz$ efficiency, the difference
of the two methods for correcting for $p^*$-dependent efficiency,
and the $\Ds\to\phi\pip$ branching fraction. The total systematic uncertainty
is calculated from the quadrature sum of all sources.
The result is the following
$\DsTT$ mass:
\begin{equation*}
m = (2319.6\pm 0.2\;(\text{stat.}) 
           \pm 1.4\;(\text{syst.}))\;\text{\mevcc} \;, 
\end{equation*}
and the following yields:
\begin{equation*}
\renewcommand{\arraystretch}{1.2}
\begin{array}{lcl@{}r@{\:}c@{\:}r@{\:}c@{\:}r@{}l}
\bar{\sigma}(\DsTT\to\Ds\piz) 
&=& (&115.8 &\pm&   2.9  &\pm&   8.7&)
\;\text{fb}  \\
\bar{\sigma}(\DsFE\to\Ds\piz) 
&=& (&-1.0 &\pm&   1.4  &\pm&    0.1&)
\;\text{fb},
\end{array}
\end{equation*}
where the first error is statistical and the second is systematic.

\begin{table}
\caption{\label{tb:dspi0.syst}A summary of systematic uncertainties
for the $\DsTT$ mass and yield from the analysis of the $\Ds\piz$ final state.
}
\begin{center}

\begin{ruledtabular}
\begin{tabular}{lrr}
       & \multicolumn{1}{c}{Mass   } &\multicolumn{1}{c}{Relative  }\\
Source & \multicolumn{1}{c}{(\mevcc)}&\multicolumn{1}{c}{yield (\%)}\\
\hline
$\Ds$ mass                &    0.6 & ---  \\
EMC energy scale          &    1.3 & ---  \\
$\DsFE$ reflection size   & $<0.1$ & 2.0  \\
$\DsFE$ mass              &    0.1 & 0.7  \\
Detector resolution       & $<0.1$ & 3.2  \\
$\DsTO$ reflection model  & $<0.1$ & 0.4  \\
$\Ds$ efficiency          &   ---  & 3.9  \\
$\piz$ efficiency         &   ---  & 3.0  \\
$p^*$ distribution        &   ---  & 0.6  \\
$\Ds\to\phi\pip$ yield
                          &   ---  & 3.9  \\
\hline
Quadrature sum            &    1.4 & 7.4  \\
\end{tabular}
\end{ruledtabular}

\end{center}
\end{table}

As can be seen in Table~\ref{tb:dspi0.syst}, the 
determination of the $\DsTT$ mass 
is limited by the understanding of the EMC energy scale. 
A more primitive calculation of this energy scale was used in
a previous estimate of the $\DsTT$ mass from this 
collaboration~\cite{Aubert:2003pe}, resulting in an 
mass estimate that is 2.3~\mevcc
lighter then the estimate presented here. The associated systematic
uncertainty in this previous work was also incorrectly calculated.
The central value and systematic uncertainty in mass reported here
reflects the current best understanding of these
calibration issues.

\begin{figure}
\begin{center}
\includegraphics[width=0.8\linewidth]{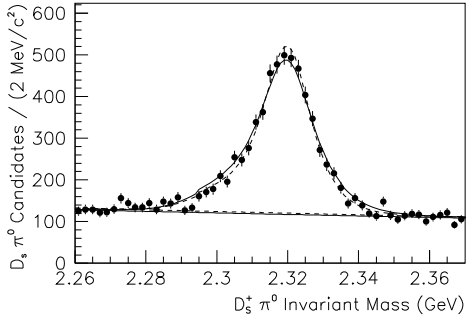}
\end{center}
\vskip -0.25in
\caption{\label{fg:dspi0.widthfigure}Likelihood fit results for 
a $\DsTT$ meson of instrinsic width 
(dashed line) $\Gamma = 0$ and (solid line)
$\Gamma = 3.8$~\mev. Shown in comparison is the mass
distribution (solid points) for the $\Ds\piz$ candidates.
}
\end{figure}

For the sake of simplicity, in order to incorporate systematic effects
into a limit on $\Gamma$, the least strict
limit obtained from the various systematic checks is quoted. This
produces a 95\% CL of $\Gamma < 3.8$~\mev. The lineshape produced
by this limit is illustrated in Fig.~\ref{fg:dspi0.widthfigure}.

\section{\boldmath The $\Ds\gamma$ Final State}

\begin{figure}
\begin{center}
\includegraphics[width=\linewidth]{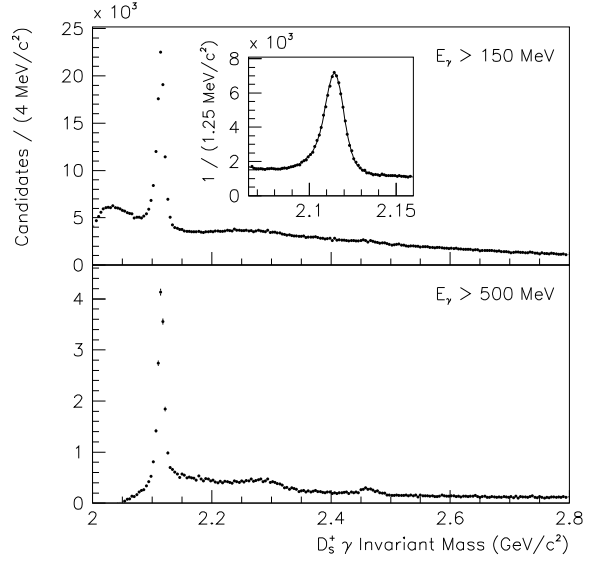}
\end{center}
\vskip -0.25in
\caption{\label{fg:dsgam.data2112paper}The invariant $\Ds\gamma$ mass
distribution using $\gamma$ candidates with loose (150~\mev) and
final (500~\mev) energy requirements.
The insert focuses on the region of the $\DsTO$ meson using the loose
energy requirement. The curve represents the fit described in the text.
}
\end{figure}

The $\Ds\gamma$ mass
distribution using $\gamma$ candidates with loose (150~\mev) and
final (500~\mev) minimum energy requirements is 
shown in Fig.~\ref{fg:dsgam.data2112paper}. 
Some structure in the vicinity of the $\DsTT$ and
$\DsFE$ masses becomes apparent once the tighter energy requirements
are applied. The looser requirement is useful for studying the
$\DsTO$ peak. A fit to that peak
consisting of two Gaussians on top of a polynomial
background function results in a peak $\DsTO$ mass of
$2113.8$~\mevcc\  and a yield of 75~000 decays, both with negligible 
statistical uncertainties.

An unbinned likelihood fit is applied
to the final $\Ds\gamma$ mass distribution in order to 
extract the parameters and
yield of the $\DsFE$ meson and upper limits on $\DsTT$ decay. 
For simplicity, the fit is performed only for masses between
2.15 and 2.85~\gevcc.
The likelihood 
fit includes five distinct sources of $\Ds\gamma$ combinations:
\begin{itemize}
\item
$\DsFE\to\Ds\gamma$ decay.
\item
$\DsTT\to\Ds\gamma$ decay (hypothetical).
\item
A reflection from $\DsTT\to\Ds\piz$ decay in which only one of the
$\gamma$ particles from $\piz$ decay is included.
\item
A reflection from $\DsFE\to\DsTO\piz$ decay in which only one of the
$\gamma$ particles from $\piz$ decay is included.
\item
Background from both unassociated $\Ds$ and $\gamma$ mesons
and the high-mass tail from $\DsTO\to\Ds\gamma$ decay.
\end{itemize}
The PDF used to describe the
mass distribution of each of these sources is described below.

\begin{figure}
\begin{center}
\includegraphics[width=0.49\linewidth]{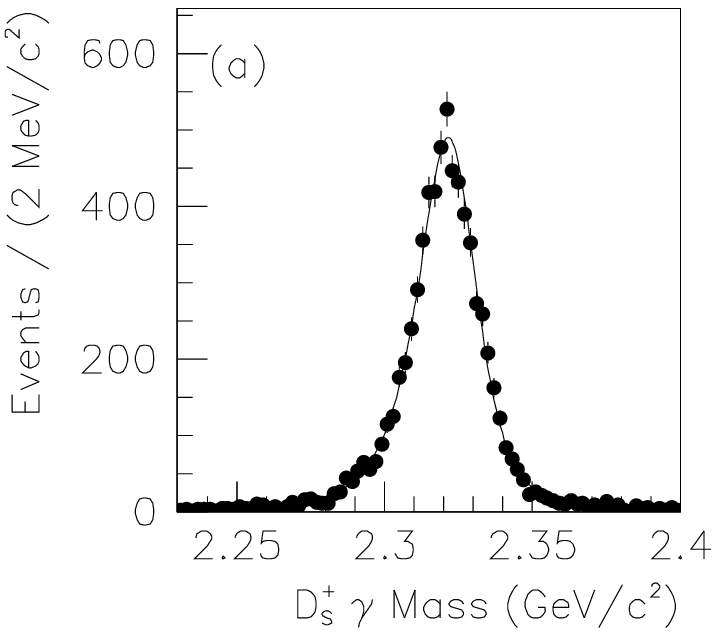}
\includegraphics[width=0.49\linewidth]{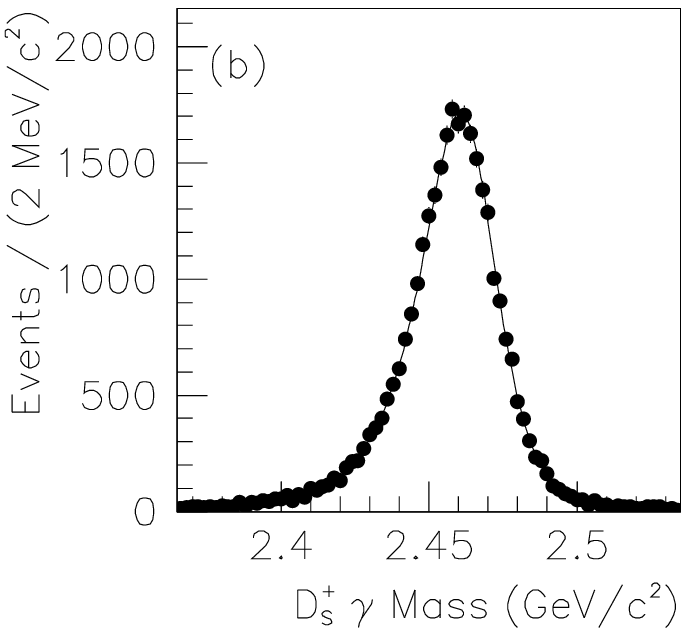}
\includegraphics[width=0.49\linewidth]{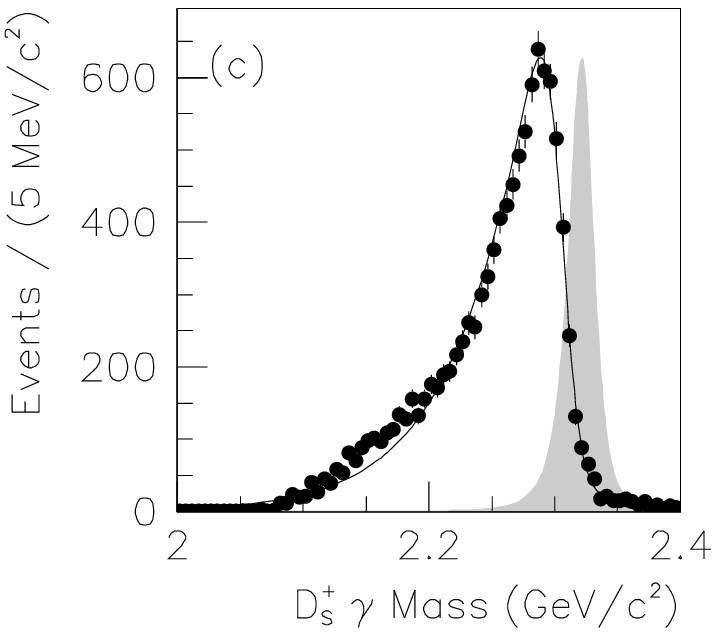}
\includegraphics[width=0.49\linewidth]{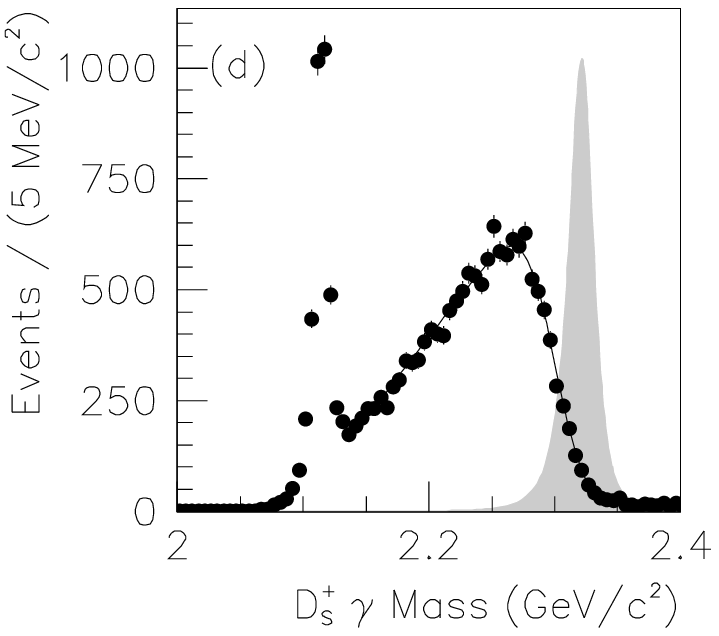}
\end{center}
\vskip -0.25in
\caption{\label{fg:dsgam.signals}The reconstructed $\Ds\gamma$ invariant mass
spectrum from MC samples for (a) $\DsTT$ and (b)
$\DsFE$ decay and (c) $\DsTT\to\Ds\piz$ and (d)
$\DsFE\to\DsTO\piz$ reflections. The curves are the fit functions
described in the text. The $\DsTT$ signal shapes from (a) and (b)
are shown for
comparison in (c) and (d) in gray.
}
\end{figure}

As shown in Fig.~\ref{fg:dsgam.signals}(b), the reconstructed
mass distribution of the $\DsFE\to\Ds\gamma$ decay, as predicted by
MC, has a long, low mass tail. To describe this 
shape, the MC sample is fit to a modified Lorentzian function
$F_{L2}(m)$:
\begin{eqnarray}
F_{L2}(m) &=& a_3 \frac{ \left| 1 + a_4 \delta  + a_5 \delta^2
                          + a_6 \delta^3 \right| }
                  { \left( 1 + \delta^2 \right)^{a_7} } 
\label{eq:modlorentz2}\\
\delta &\equiv& \left( m - a_1 \right)/a_2 \;. \nonumber
\end{eqnarray}                  
The fit results are shown in Fig.~\ref{fg:dsgam.signals}(b).
A similar procedure is used for the hypothetical $\DsTT\to\Ds\gamma$
decay (Fig.~\ref{fg:dsgam.signals}(a)).

Ignoring resolution effects,
the $\DsTT$ reflection produces an invariant $\Ds\gamma$ mass
distribution up to a maximum of approximately
140~\mevcc\  below the $\DsTT$ mass. Candidate selection requirements
produce a distribution that peaks at this limit. Resolution effects
smear this sharp peak producing the shape shown
in Fig.~\ref{fg:dsgam.signals}(c), as predicted by MC.
This distribution can be reasonably described (in the mass range
of interest) by a bounded quadratic
function convoluted with a double Gaussian. The $\DsFE$ reflection
has a similar behavior near the $\DsTT$ mass (Fig.~\ref{fg:dsgam.signals}(d)).
Both distributions overlap the direct $\DsTT\to\Ds\gamma$ decay.

The following function $D(m)$ is used to represent the remainder of 
the $\Ds\gamma$ distribution:
\begin{equation}
D(m) = 1 + a_1 \exp\left[ -\right(m - a_3\left)^2/a_2 \right] \;.
\label{eq:dsgam.background}
\end{equation}
This includes combinatorial background along with any tail from 
$\DsTO\to\Ds\gamma$ decay. The MC simulation fails to reproduce the
shape of this background, either due to unknown $c\bar{s}$ contributions
(for example, higher resonant states) or unexpected behavior of the
tail of the distribution from $\DsTO$ decay. This issue, combined
with the complex shapes associated with the $\DsTT$ and $\DsFE$ reflections,
leads to considerable
systematic uncertainty in the fit. Likelihood fits under several different
conditions are attempted in order to understand the uncertainty. 

\begin{figure}
\vskip -0.15in
\includegraphics[width=\linewidth]{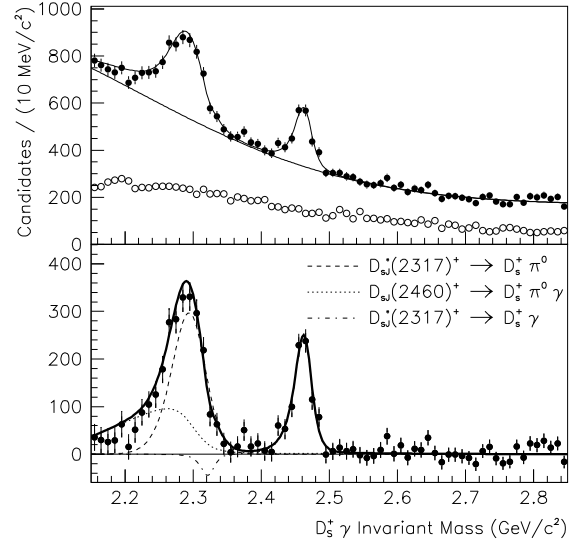}
\vskip -0.15in
\caption{\label{fg:dsgam.plotfitdatapaper}An example likelihood fit
to the $\Ds\gamma$ invariant mass distribution. The solid points in the top
plot are the mass distribution. The open points are the $\Ds$ sidebands,
scaled appropriately. The bottom plot shows the same data 
after subtracting the background curve from the fit. Various contributions
to the likelihood fit are also shown.
}
\end{figure}

One fit that produces a good representation of the data is shown in
Fig.~\ref{fg:dsgam.plotfitdatapaper}. In this fit, all parameters 
except the upper mass limit of the $\DsTT$ reflection are allowed to
vary. The estimated raw $\DsFE$ ($\DsTT$)
yield from this fit is $920 \pm 60$ ($-130 \pm 130$).
The fitted $\DsFE$ mass is $(2459.5\pm 1.2)$~\mevcc\ (statistical
errors only).

\begin{figure}
\begin{center}
\includegraphics[width=0.8\linewidth]{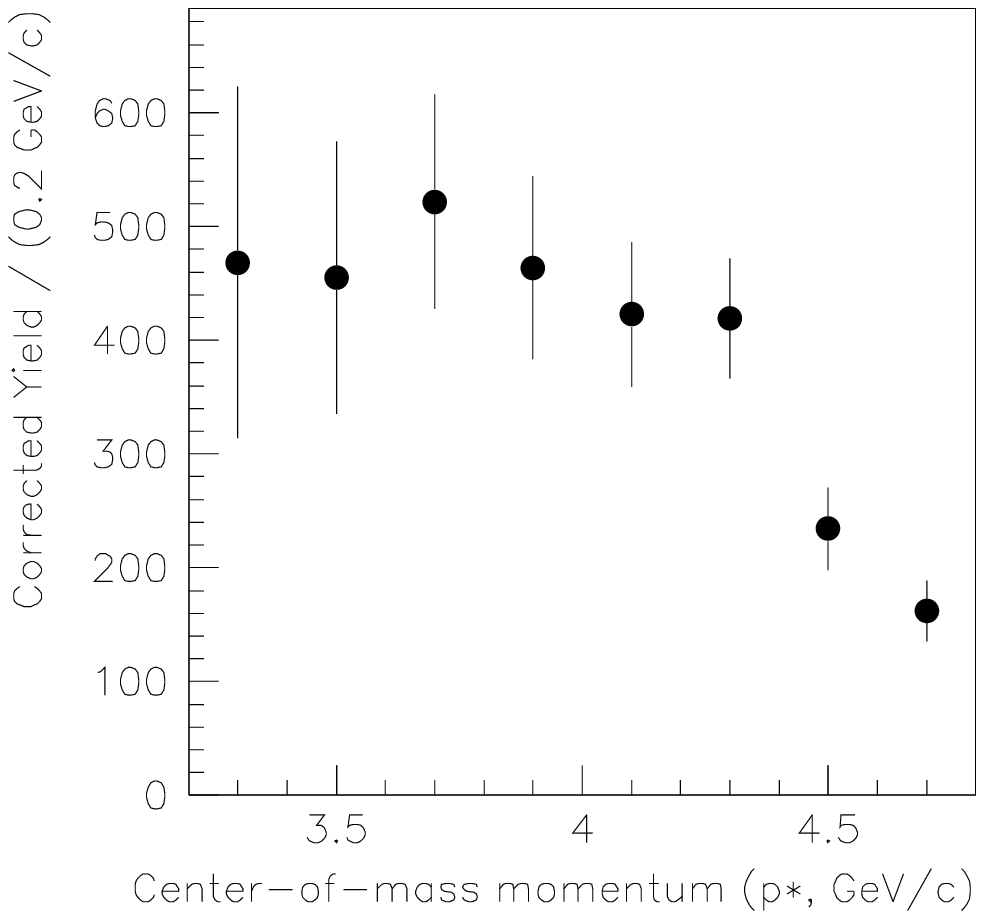}
\end{center}
\vskip -0.25in
\caption{\label{fg:dsgam.pstarpaper}Corrected $\DsFE\to\Ds\gamma$ yield
as a function of $p^*$.
}
\end{figure}

The $\DsFE$ signal is far enough removed from the $\DsTT$ and $\DsFE$
reflections that accurate mass and yield results are obtained.
The same two methods described in the previous section are used to
estimate the $\DsFE$ yield. The first method, using $p^*$ dependent
weights proportional to the inverse of efficiency, produces a corrected
yield of:
\begin{equation*}
N(\DsFE\to\Ds\gamma,\Ds\to\phi\pip) = 3270 \pm 230
\end{equation*}
for $p^* > 3.2$~\gevc\ (statistical error only). The second method
produces the $p^*$ spectrum shown in Fig.~\ref{fg:dsgam.pstarpaper}.
The total yield from the spectrum is $3080 \pm 240$ (statistical error only),
approximately 6.2\% lower than the first estimate.

Systematic uncertainties associated with the
assumed PDFs for the background and reflection shapes are explored using
different variations of the likelihood fit.
Among the variations
applied is an alternate description of the background shape:
\begin{equation}
D'(m) = 1 + a_1 \exp\left[ -5\right(m - a_3\left) \right] 
   + a_2 / \left(m - a_3\right)^2 \;.
\label{eq:dsgam.altbackground}
\end{equation}
In addition, the MC predictions for the size and shape
of the $\DsTT$ reflection are used unaltered (despite producing a fit
of inferior quality). Large variations of raw $\DsTT$ yield of
up to 490 events are observed. 

The shape of the $\DsFE\to\Ds\gamma$ signal is sensitive to several
factors that are difficult to simulate exactly, including
EMC energy resolution. Variations in the assumed resolution are
used to study the associated systematic uncertainty in yield and mass.
The result is an uncertainty of 3.5\% in  yield and
no significant change in mass.

All systematic uncertainties
for the $\DsFE$ mass and yield are listed in 
Table~\ref{tb:dsgam.syst}. The result is the following
$\DsFE$ mass:
\begin{equation*}
m = (2459.5\pm 1.2\;(\text{stat.}) 
           \pm 3.7\;(\text{syst.}))\;\text{\mevcc} \;, 
\end{equation*}
and the following yields:
\begin{equation*}
\renewcommand{\arraystretch}{1.2}
\begin{array}{lcl@{}r@{\:}c@{\:}r@{\:}c@{\:}r@{}l}
\bar{\sigma}(\DsTT\to\Ds\gamma) 
&=& (&-2.4 &\pm&   2.3  &\pm&   8.9&)
\;\text{fb}  \\
\bar{\sigma}(\DsFE\to\Ds\gamma) 
&=& (&14.4 &\pm&   1.0  &\pm&   1.4&)
\;\text{fb},
\end{array}
\end{equation*}
where the first error is statistical and the second is systematic.

\begin{table}
\caption{\label{tb:dsgam.syst}A summary of systematic uncertainties
for the $\DsFE$ mass and yield from the analysis of the $\Ds\gamma$ final state.
}
\begin{center}

\begin{ruledtabular}
\begin{tabular}{lrrr}
       & \multicolumn{1}{c}{Mass   } &\multicolumn{1}{c}{Relative }\\
Source & \multicolumn{1}{c}{(\mevcc)}&\multicolumn{1}{c}{yield (\%)}\\
\hline
$\Ds$ mass                &    0.6 & ---     \\
EMC energy scale          &    3.7 & ---     \\
$\DsTT$ mass              &    0.1 & $ 0.1$  \\
$\DsTT$ reflection shape  &    0.1 & $ 1.9$  \\
Detector resolution       & $<0.1$ & $ 3.5$  \\
Background shape          &    0.5 & $ 3.3$  \\
$\Ds$ efficiency          &   ---  & $ 3.9$  \\
$\gamma$ efficiency       & $<0.1$ & $ 1.8$  \\
$p^*$ distribution        &   ---  & $ 6.3$  \\
$\Ds\to\phi\pip$ branching fraction
                          &   ---  & $ 3.9$  \\
\hline                    
Quadrature sum            &   3.7  & $10.0$  \\
\end{tabular}
\end{ruledtabular}

\end{center}
\end{table}

\section{\boldmath The $\Ds\piz\gamma$ Final State}
\label{sec:dspi0gam}

\begin{figure}
\begin{center}
\includegraphics[width=0.49\linewidth]{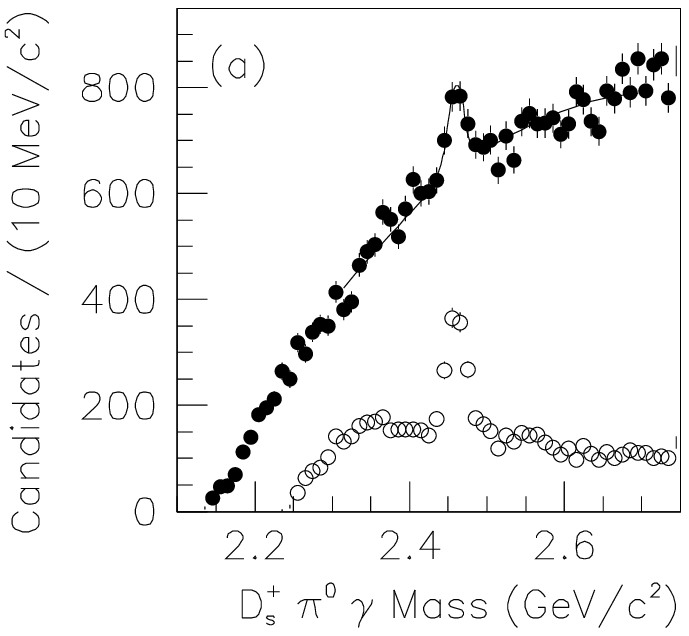}
\includegraphics[width=0.49\linewidth]{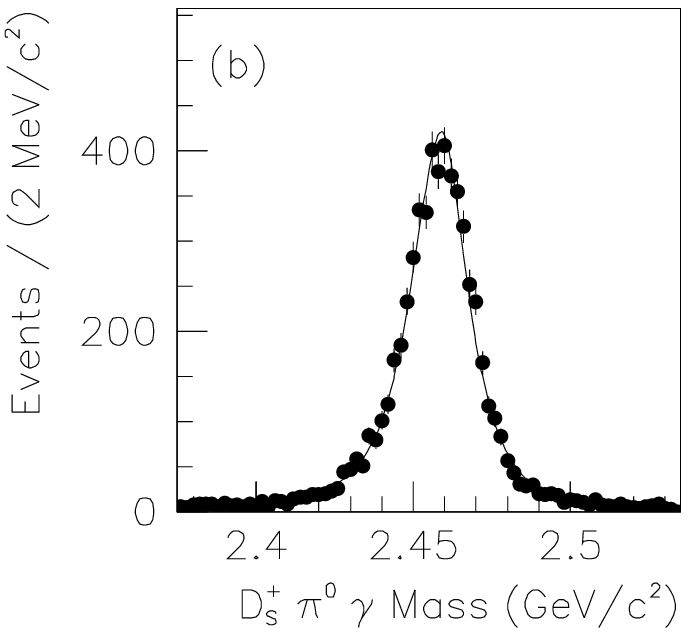}
\end{center}
\vskip -0.25in
\caption{\label{fg:dspi0gam.dataallpaper}(a) The
sample of $\Ds\piz\gamma$ candidates shown in solid points. 
The fit described in the text is overlaid. 
The open points are those candidates which fall in a restricted
$\Ds\gamma$ mass range.
(b) The reconstructed $\Ds\piz\gamma$ invariant mass
spectrum from a $\DsFE\to\DsTO\piz$ MC sample.
The fit function of Eq.~\ref{eq:modlorentz3} is overlaid.}
\end{figure}

The 
invariant mass spectrum for all selected  $\Ds\piz\gamma$ candidates
is shown in Fig.~\ref{fg:dspi0gam.dataallpaper}(a).
A $\DsFE$ signal is apparent. The shape of this signal is characterized by 
applying the following modified Lorentzian fit function $F_{L3}$:
\begin{eqnarray}
F_{L3}(m) &=& a_3 \frac{ \left| 1 + a_4 \delta  
                          + a_5 \tan^{-1} 5\delta \right| }
                  { \left( 1 + \delta^2 \right)^{a_6} } 
\label{eq:modlorentz3}\\
\delta &\equiv& \left( m - a_1 \right)/a_2 \;. \nonumber
\end{eqnarray}             
to a $\DsFE\to\DsTO\piz$ MC sample (Fig.~\ref{fg:dspi0gam.dataallpaper}(b)).
A binned $\chi^2$ fit to the $\Ds\piz\gamma$ spectrum that includes
this shape along with a polynomial description of the background
produces a $\DsFE$ mass $(2459.5 \pm 2.0)$~\mevcc\  and a yield of
$560 \pm 80$ events (statistical errors only).

\begin{figure}
\begin{center}
\includegraphics[width=0.8\linewidth]{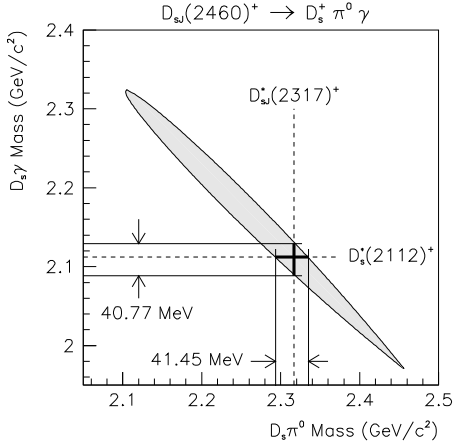}
\end{center}
\vskip -0.25in
\caption{\label{fg:dspi0gam.dalitzpaper}The light gray region 
indicates the range of $\Ds\piz$ and $\Ds\gamma$
mass that is kinematically allowed in the decay of an object of mass
$2458.0$~\mevcc\ to $\Ds\piz\gamma$. 
The lines mark the kinematic space associated with
decays which proceed through an intermediate $\DsTO$ or $\DsTT$ meson. 
}
\end{figure}

In the following, it is assumed 
that the $\DsFE$ meson decays to $\Ds\piz\gamma$
entirely through either of the
two kinematically allowed sub-resonant decay modes:
\begin{equation}
\DsFE \to
\left\{
   \begin{array}{c}
   \DsTO\piz \\
   \DsTT\gamma \
   \end{array}
\right\} \to \Ds\piz\gamma
\label{eq:dspi0gam.submodes}
\end{equation}
Due to a kinematic accident, the phase space of these two sub-resonant
modes overlap, as illustrated in 
Fig.~\ref{fg:dspi0gam.dalitzpaper}. It is therefore possible to
remove background while retaining both sub-resonant decay modes
by selecting $\Ds\piz\gamma$ candidates
in either a restricted range of $\Ds\piz$ mass or a restricted
range of $\Ds\gamma$ mass.
This analysis uses a requirement that the $\Ds\gamma$ mass
must be within 20~\mevcc\  of the $\DsTO$ mass. As shown in 
Fig.~\ref{fg:dspi0gam.dataallpaper}(a), the resulting $\Ds\piz\gamma$
mass distribution in this $\Ds\gamma$ signal window is considerably cleaner.

The $\Ds\gamma$ requirement introduces a source of background
that peaks underneath the $\DsFE$ signal. This background is a reflection
from $\DsTT\to\Ds\piz$ decays that are not associated with any $\DsFE$ decay. 
The reflection arises because,
as illustrated in Fig.~\ref{fg:dspi0gam.dalitzpaper},
any $\DsTT$ signal that is combined with a $\gamma$ candidate that produces
a $\Ds\gamma$ mass near the $\DsTO$ meson results in
a $\Ds\piz\gamma$ mass near the $\DsFE$ meson. 
If the $\Ds\gamma$ mass requirement is shifted upwards or downwards, 
this $\DsTT$ reflection shifts up and down in
$\Ds\piz\gamma$ mass by a predictable amount.

Another type of background that behaves similarly to the $\DsTT$
reflection is $\DsFE\to\Ds\piz\gamma$ decay in which the wrong $\gamma$
candidate is chosen. This type of background is slightly wider and smaller
than the $\DsTT$ reflection, but otherwise has a similar 
mass distribution. To describe this background
contribution, it is assumed that the $\DsFE$ decays entirely through
$\DsTO\piz$ and is produced at a rate comparable to previous measurements.
Both assumptions need not be entirely accurate since this background
has a relatively small contribution.

\begin{figure}
\begin{center}
\includegraphics[width=0.8\linewidth]{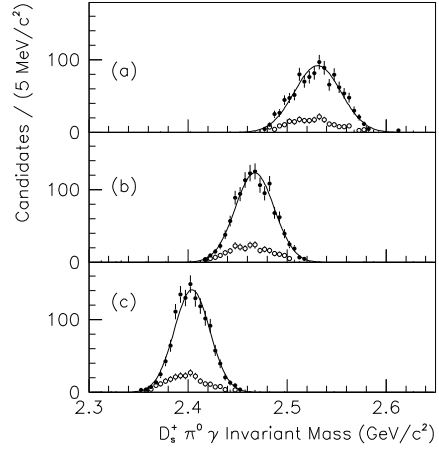}
\end{center}
\vskip -0.25in
\caption{\label{fg:dspi0gam.signalbackpaper}The combined $\Ds\piz\gamma$
invariant mass distribution (solid points) as obtained from MC samples
for the combination of the $\DsTT$ reflection 
and $\DsFE\to\DsTO\piz$ decays in which the incorrect $\gamma$
is chosen. The $\DsFE$ contribution alone is shown in open points.
Shown are the (a) upper, (b) signal,
and (c) lower $\Ds\gamma$ mass selection windows. The curves are
fits to Gaussian distributions.
}
\end{figure}

To characterize the $\DsTT$ and $\DsFE$ reflections, upper and
lower $\Ds\gamma$ mass selection windows are chosen
centered at $\pm 60$~\mevcc\  away from the $\DsTO$ mass.
The mass distribution from MC samples of $\DsTT\to\Ds\piz$ 
and $\DsFE\to\DsTO\gamma$ decay
are shown in Fig.~\ref{fg:dspi0gam.signalbackpaper} for the signal
and two sideband $\Ds\gamma$ mass windows. The shape of the two
combined reflections
in all three cases can be successfully described by a fit to a 
Gaussian.

To determine the mass, width, and yield of the $\DsFE$ meson,
an unbinned likelihood fit is
applied to the $\Ds\piz\gamma$ mass distribution of candidates
selected in the $\Ds\gamma$ signal window. This fit includes
the following contributions:
\begin{itemize}
\item
$\DsFE\to\Ds\piz\gamma$ decay.
\item
The combined reflections from $\DsTT\to\Ds\piz$ decay
and from $\DsFE\to\DsTO\gamma$ decay in which the incorrect $\gamma$
candidate is chosen.
\item
A reflection from $\DsTO\to\Ds\gamma$ decay in which an unassociated
$\gamma$ candidate is added to form a fake $\piz$ candidate.
\item
Smooth background sources that do not have any peaking behavior.
\end{itemize}
The $\DsTO\to\Ds\gamma$ reflection is similar to that observed
in $\Ds\piz$ combinations 
(see, for example, Fig.~\ref{fg:dspi0.paperfitfig2}).
The smooth background is represented by the $C(m)$ function
described in Eq.~\ref{eq:cthres}.

\begin{figure}
\begin{center}
\includegraphics[width=\linewidth]{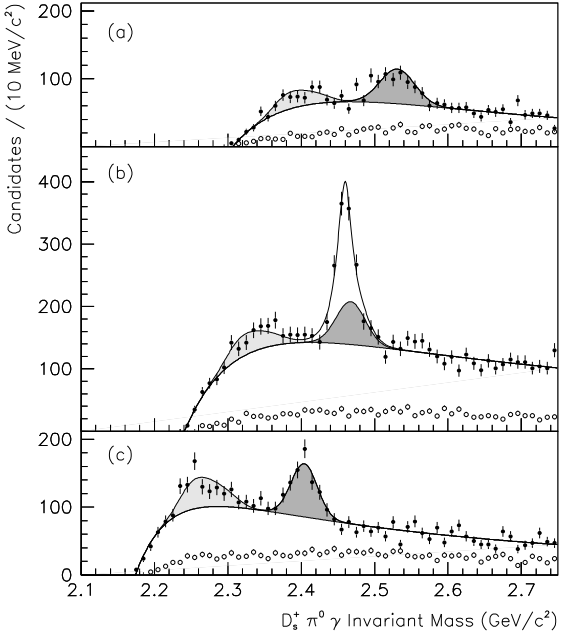}
\end{center}
\vskip -0.25in
\caption{\label{fg:dspi0gam.plotfitpaper}The invariant mass distribution
of $\Ds\piz\gamma$ candidates in the (a) upper, (b) signal,
and (c) lower $\Ds\gamma$ mass selection windows for (solid points)
the $\Ds$ signal and (open points) $\Ds$ sideband samples. The curves
represent the fits described in the text. The dark gray (light gray)
region corresponds to the predicted contribution from the
$\DsTT$ ($\DsTO$) reflection.
}
\end{figure}

Two similar fits excluding the $\DsFE$ signal
are applied to the upper and lower $\Ds\gamma$ mass samples.
These fits suggest that the MC prediction of the absolute rate of
the $\DsTO\to\Ds\gamma$ reflection is approximately 21\% too low.
The fit models are adjusted accordingly. The fit results for
all three $\Ds\gamma$ mass ranges after this adjustment are shown in
Fig.~\ref{fg:dspi0gam.plotfitpaper}. The result is
a $\DsFE$ mass of $(2458.6\pm 1.3)$~\mevcc\  and a raw
yield of $560\pm 40$ events (statistical errors only).

Although the overall size of the $\DsTT\to\Ds\piz$ reflection
is allowed to vary, the MC prediction for the relative contributions
in the three $\Ds\gamma$ mass windows is preserved. Since the size
of this reflection is adequately modeled in the two $\Ds\gamma$
sidebands, there is some confidence that the size is well
established in the $\Ds\gamma$ signal window.

Note that the size of the smooth background is relatively larger
in the signal $\Ds\gamma$ window due to contributions from
$\DsTO\to\Ds\gamma$ decay. In addition, if there were significant
non-resonant contributions to $\DsFE$ decay, peaks at the
$\DsFE$ mass would be visible in the two $\Ds\gamma$ sidebands.
No such evidence is visible.

\begin{figure}
\begin{center}
\includegraphics[width=0.8\linewidth]{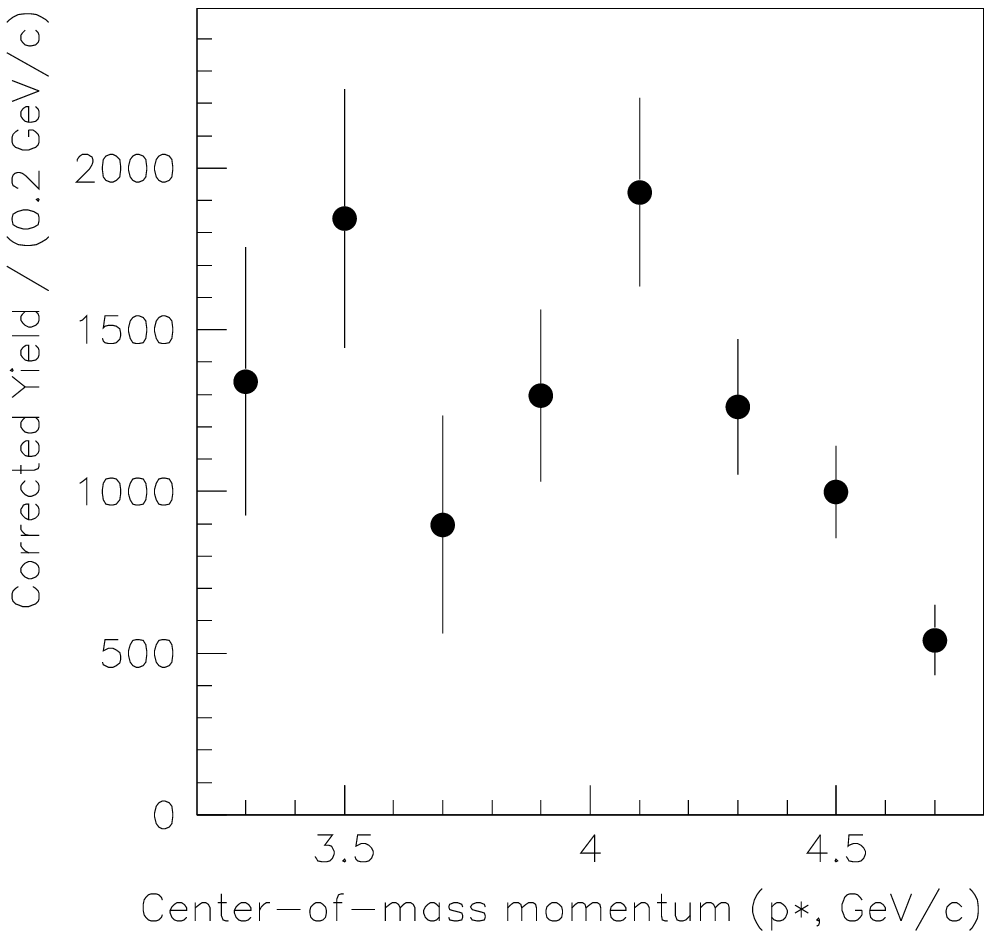}
\end{center}
\vskip -0.25in
\caption{\label{fg:dspi0gam.pstarpaper}Corrected $\DsFE\to\Ds\piz\gamma$ yield
as a function of $p^*$.
}
\end{figure}

The two methods described in the two previous sections are used to
estimate the $\DsFE$ yield. The first method, using $p^*$ dependent
weights proportional to the inverse of efficiency, produces a corrected
yield of:
\begin{equation*}
N(\DsFE\to\Ds\piz\gamma,\Ds\to\phi\pip) = 9690 \pm 790 
\end{equation*}
for $p^* > 3.2$~\gevc\ (statistical error only). The second method
produces the $p^*$ spectrum shown in Fig.~\ref{fg:dspi0gam.pstarpaper}.
The total yield from the spectrum is $9890 \pm 810$ (statistical error only).

The systematic uncertainties in the mass and yield of the
$\DsFE$ meson are shown in Table~\ref{tb:dspi0gam.syst}. 
As described previously, 
the size of the $\DsTT$ reflection was adjusted in the fit to match the
$\Ds\gamma$ sideband samples. If the size of the reflection
is taken unchanged from MC predictions, the $\DsFE$ yield increases by 7.3\%.
A second likelihood fit described later in this section 
used to distinguish
between the two $\DsFE$ sub-resonant decay modes also produces an estimate
of $\DsFE$ yield. The difference between the two fits
is treated as a systematic uncertainty.
The total $\DsFE$ yield, without distinguishing between the
two possible sub-resonant decay modes, is measured to be
\begin{equation*}
\bar{\sigma}(\DsFE\to\Ds\piz\gamma) =
(42.7 \pm 3.5 \pm 4.2)\;\text{fb},
\end{equation*}
where the first error is statistical and the second is systematic.
The complete $\DsFE$ mass result is:
\begin{equation*}
m = (2458.6\pm 1.0\;(\text{stat.}) 
           \pm 2.5\;(\text{syst.}))\;\text{\mevcc} \;. 
\end{equation*}

\begin{table}
\caption{\label{tb:dspi0gam.syst}A summary of systematic uncertainties
for the $\DsFE$ mass and yield from the analysis of the 
$\Ds\piz\gamma$ final state.
}
\begin{center}

\begin{ruledtabular}
\begin{tabular}{lrr}
       & \multicolumn{1}{c}{Mass   } &\multicolumn{1}{c}{Relative  }\\
Source & \multicolumn{1}{c}{(\mevcc)}&\multicolumn{1}{c}{yield (\%)}\\
\hline
$\Ds$ mass                &    0.6 &  ---   \\
EMC energy scale          &    2.4 &  ---   \\
$\DsTT$ reflection size   &    0.3 &  7.3   \\
$\DsTT$ mass              &    0.1 &  1.2   \\
Detector resolution       & $<0.1$ &  2.1   \\
$\DsTO$ reflection model  &   ---  &  ---   \\

Fit method                &   ---  &  1.4   \\
$\Ds$ efficiency          &   ---  &  3.9  \\
$\piz$ and $\gamma$ efficiency
                          &   ---  &  3.0   \\
$p^*$ distribution        &   ---  &  2.1   \\
$\Ds\to\phi\pip$ yield
                          &   ---  &  3.9   \\
\hline
Quadrature sum            &    2.5 &  10.2   \\
\end{tabular}
\end{ruledtabular}

\end{center}
\end{table}

The $\DsFE$ signal PDF used in the likelihood fit of 
Fig.~\ref{fg:dspi0gam.plotfitpaper} includes a $\DsFE$ intrinsic width
of $\Gamma = 0.1$~\mev. Larger intrinsic widths do not result in any
significant improvement of the fit.
After applying the same likelihood-integration technique
described in Section~\ref{sec:dspi0} for the $\Ds\piz$ final state,
and including the systematic effects listed in 
Table~\ref{tb:dspi0gam.syst},
the result is a 95\% CL limit of $\Gamma<6.3$~\mev.

\begin{figure}
\begin{center}
\includegraphics[width=0.49\linewidth]{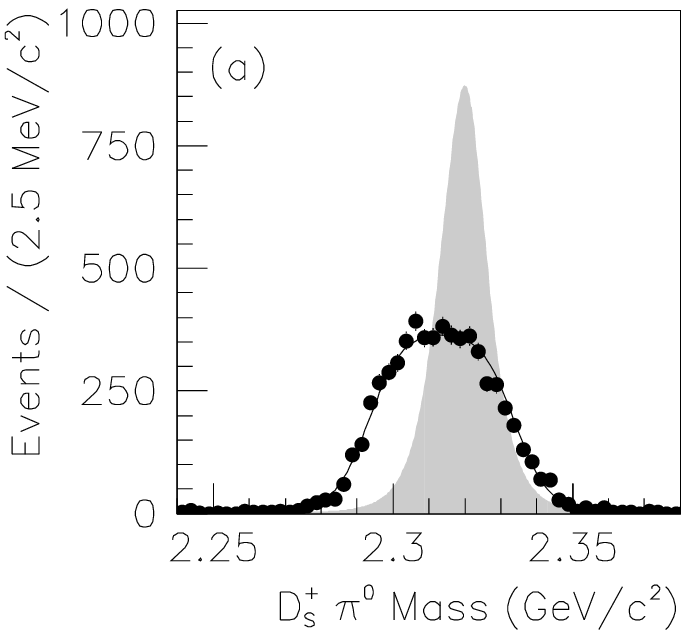}
\includegraphics[width=0.49\linewidth]{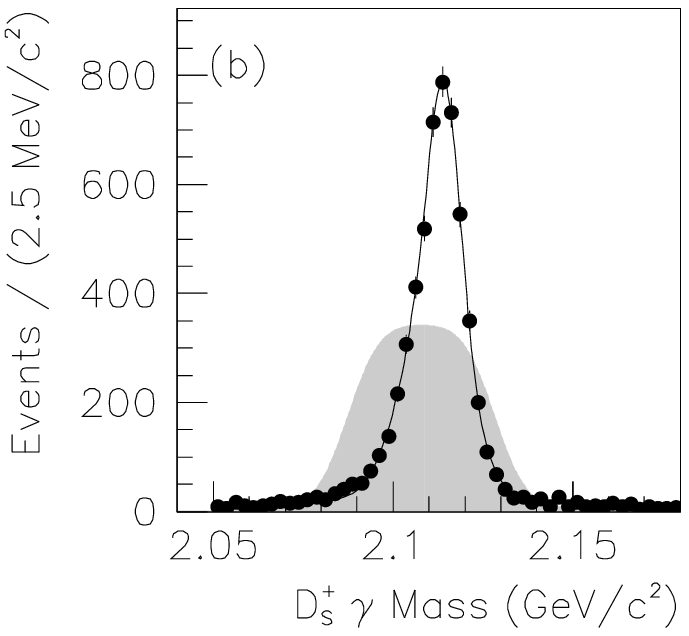}
\end{center}
\vskip -0.25in
\caption{\label{fg:dspi0gam.signals}The 
reconstructed 
(a) $\Ds\piz$ and (b) $\Ds\gamma$ invariant mass distributions for the two
possible $\DsFE$ sub-resonant decay modes. The
distributions from a $\DsFE\to\DsTO\piz$ MC sample are shown in points. 
The curves are the
fits described in the text. The shaded regions are the shapes
assumed for $\DsFE\to\DsTT\gamma$  decay.
}
\end{figure}

Having established a $\DsFE\to\Ds\piz\gamma$ signal, it is now
necessary to distinguish between the two possible sub-resonant decay modes
shown in Eq.~\ref{eq:dspi0gam.submodes}. These two decay modes 
can be distinguished
by their $\Ds\piz$ and $\Ds\gamma$ invariant mass distributions,
as shown in Fig.~\ref{fg:dspi0gam.signals}.
The distributions for the $\DsFE\to\DsTO\piz$ subresonant mode
are determined using a MC sample. The reconstructed
$\Ds\gamma$ mass distribution,
which is relatively narrow (as it arises from $\DsTO$ decay),
is represented by a $\chi^2$ fit to the $F_{L3}$ function 
(Eq.~\ref{eq:modlorentz3}).
The wider $\Ds\piz$ mass distribution is accurately modeled by a
square function smeared by a double Gaussian. Both fits are shown
in Fig.~\ref{fg:dspi0gam.signals}.

In contrast, for the $\DsFE\to\DsTT\gamma$ subresonant decay mode,
the $\Ds\piz$ mass distribution is narrow and the $\Ds\gamma$ mass
distribution is wide. The $\Ds\piz$ distribution is determined using a
$\DsTT\to\Ds\piz$ MC sample. The $\Ds\gamma$ distribution is
calculated using the parameters determined from the $\Ds\piz$
distribution from $\DsFE\to\DsTO\piz$ described above converted to the
appropriate kinematic range. The shapes assumed for both
$\DsFE\to\DsTT\gamma$ mass projections are shown in gray in
Fig.~\ref{fg:dspi0gam.signals}.

\begin{figure}
\begin{center}
\includegraphics[width=0.6\linewidth]{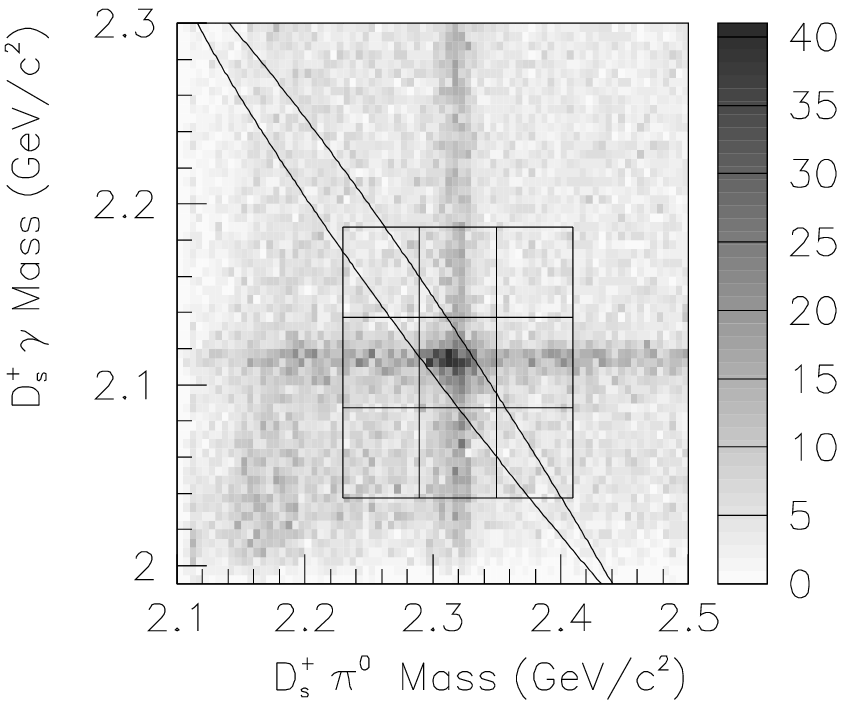}
\end{center}
\vskip -0.25in
\caption{\label{fg:dspi0gam.rangepapernew}The $\Ds\gamma$ versus
$\Ds\piz$ mass distributions for the $\Ds\piz\gamma$ candidates.
The horizontal (vertical) band corresponds to background from
$\DsTO\to\Ds\gamma$ ($\DsTT\to\Ds\piz$) decay. The excess of candidates
near the crossing of these two bands is the $\DsFE$ signal.
The curve indicates the region of phase space in which the $\DsFE$ decay
is kinematically restricted.
The grid identifies the subsample of candidates used in the likelihood
fit shown in Fig.~\ref{fg:dspi0gam.fitoverplotpaper}.
}
\end{figure}

The $\Ds\piz$ and $\Ds\gamma$ mass distributions of the signal
cannot be explored without correctly subtracting backgrounds
from unassociated $\DsTO\to\Ds\gamma$ and $\DsTT\to\Ds\piz$ decay.
This
subtraction is performed by a two-dimensional unbinned likelihood fit applied
to the $\Ds\piz$ and $\Ds\gamma$ mass distributions of the data.
The likelihood fit is restricted to the data sample contained inside
the grid shown
in Fig.~\ref{fg:dspi0gam.rangepapernew}. This fit includes five sources of
$\Ds\piz\gamma$ candidates:
\begin{itemize}
\item
Combinatorial background represented by a two-dimensional quadratic
function.
\item
Background from $\DsTO\to\Ds\gamma$ decay combined with unassociated
$\piz$ candidates represented by a $\DsTO$ line shape in the $\Ds\gamma$
mass and as a linear function in $\Ds\piz$ mass.
\item
Background from $\DsTT\to\Ds\piz$ decay combined with unassociated
$\gamma$ candidates represented by a $\DsTT$ line shape in the $\Ds\piz$
mass and as a linear function in $\Ds\gamma$ mass.
\item
A signal from $\DsFE\to\DsTO\piz$ with $\Ds\piz$ and
$\Ds\gamma$ mass distributions represented by the curves
in Fig.~\ref{fg:dspi0gam.signals}.
\item
A signal from $\DsFE\to\DsTT\gamma$ with $\Ds\piz$ and
$\Ds\gamma$ mass distributions represented by the gray regions 
in Fig.~\ref{fg:dspi0gam.signals}.
\end{itemize}

\begin{figure}
\begin{center}
\includegraphics[width=\linewidth]{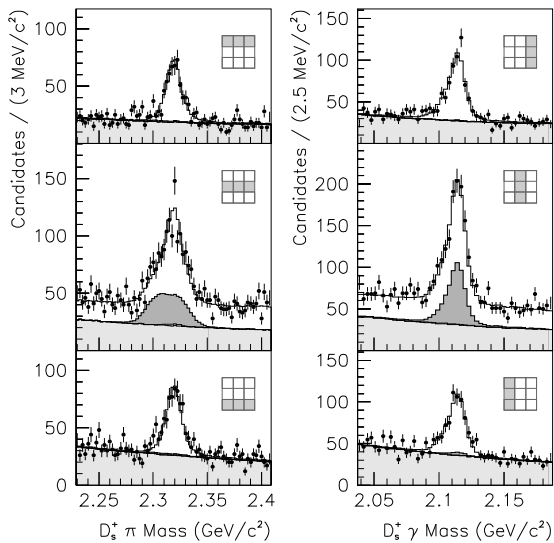}
\end{center}
\vskip -0.25in
\caption{\label{fg:dspi0gam.fitoverplotpaper}The $\Ds\piz$ and $\Ds\gamma$
invariant mass distributions for $\Ds\piz\gamma$ candidates that fall
within the indicated portions of the 
grid shown in Fig.~\ref{fg:dspi0gam.rangepapernew}. The
histograms represent the results of a likelihood fit. The light gray
region corresponds to combinatorial background. The medium gray
(dark gray) region represents the fitted fraction of
$\DsFE\to\DsTO\piz$ ($\DsFE\to\DsTT\gamma$).
}
\end{figure}

The result of this likelihood fit is shown in 
Fig.~\ref{fg:dspi0gam.fitoverplotpaper},
divided into the regions delineated by the grid shown in
Fig.~\ref{fg:dspi0gam.rangepapernew}. The fit produces an adequate model of
the data in all regions. The result (statistical errors only)
is a total yield
of $520\pm 50$ $\DsFE\to\Ds\piz\gamma$ decays with a 
fraction of ($2.5\pm 8.8$)\% 
proceeding through the $\DsTT\gamma$ channel,
the former number being somewhat smaller than the yield determined by
the $\Ds\piz\gamma$ mass fit (Fig.~\ref{fg:dspi0gam.plotfitpaper}), 
though consistent within systematic uncertainties.
Based on these
results, it appears that the decay $\DsFE\to\Ds\piz\gamma$ can be
successfully 
described as proceeding entirely through the channel $\DsTO\piz$.

The systematic uncertainties listed in Table~\ref{tb:dspi0gam.syst}
can be applied to the above fit results. In combination with the
results of the fit to the $\Ds\piz\gamma$ mass distribution
of Fig.~\ref{fg:dspi0gam.plotfitpaper}, and treating correlated
systematic uncertainties in the appropriate fashion, the following
yields for the subresonant specific decays are obtained:
\begin{equation*}
\renewcommand{\arraystretch}{1.2}
\begin{array}{lcl@{}r@{\:}c@{\:}r@{\:}c@{\:}r@{}l}
\bar{\sigma}(\DsFE\to\DsTO\piz) 
&=& (&41.6 &\pm&   5.1  &\pm&   5.0&)
\;\text{fb}  \\
\bar{\sigma}(\DsFE\to\DsTT\gamma) 
&=& (& 1.1 &\pm&   5.1  &\pm&   5.0&)
\;\text{fb},
\end{array}
\end{equation*}
where the first error is statistical and the second is systematic.

A simple helicity analysis
is used to test the $J^P$ assignment of the $\DsFE$ meson
under the assumption that the decay $\DsFE\to\Ds\piz\gamma$ proceeds entirely
through the subresonant mode $\DsTO\piz$.
This analysis is performed in terms of the helicity angle $\vartheta_h$,
defined as the angle of the $\gamma$ in the $\DsTO$ center-of-mass
frame with respect to the $\DsTO$ direction. Since the $\DsTO$ is
a vector particle, the helicity distribution must consist of some combination
of the zero helicity distribution $H_0$:
\begin{equation}
H_0 = \sin^2\vartheta_h
\end{equation}
and the helicity = $\pm 1$ distribution $H_1$:
\begin{equation}
H_1 = \frac{1}{2}\left( 1 + \cos^2\vartheta_h \right)\;.
\end{equation}
As listed in Table~\ref{tb:dspi0gam.helicity},
the expected combination of $H_0$ and $H_1$ depends on the assumed
$\DsFE$ spin and parity.

\begin{table}
\caption{\label{tb:dspi0gam.helicity}Helicity distributions 
in the decay $\DsFE\to\DsTO\piz$ for 
hypothetical $\DsFE$ spin-parity assignments.
}
\begin{center}

\begin{ruledtabular}
\begin{tabular}{ll}
$J^P$ & Helicity Distribution \\
\hline
$0^+$ & (decay is forbidden) \\
$0^-$ & $H_0$ \\
$1^-,2^+,3^-,4^+,\ldots$ & $H_1$ \\
$1^+,2^-,3^+,4^-,\ldots$ & any combination of $H_0$ and $H_1$ \\
\end{tabular}
\end{ruledtabular}

\end{center}
\end{table}

\begin{figure}
\begin{center}
\includegraphics[width=0.6\linewidth]{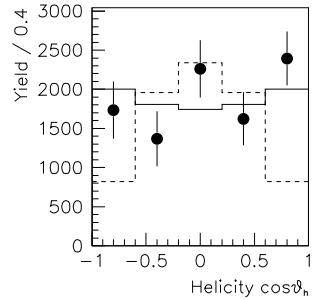}
\end{center}
\vskip -0.25in
\caption{\label{fg:dspi0gam.helicity}Efficiency-corrected yields in
five $\cos\vartheta_h$ bins for the decay $\DsFE\to\DsTO\piz$. 
The solid histogram is the result of a fit to
the function described in the text. The dashed histogram is a similar
fit with the $a_2$ parameter fixed to zero.
}
\end{figure}

To measure the helicity distribution, the $\Ds\piz\gamma$ candidates
are divided into five bins of $\cos\vartheta_h$. The $\Ds\piz\gamma$ mass
fit of Fig.~\ref{fg:dspi0gam.plotfitpaper} is repeated for each of these
subsamples using $p^*$-dependent weights inversely proportional 
to the selection
efficiency in order to correct for acceptance. The result is shown
in Fig.~\ref{fg:dspi0gam.helicity}. The integral of the following
function is calculated in each $\cos\vartheta_h$ bin:
\begin{equation}
F_h = a_1 \left[ \left( 1 - a_2 \right) H_0 + a_2 H_1 \right] \;.
\end{equation}
A $\chi^2$ fit is used to determine the most likely value of $a_2$. The
result is $a_2 = 0.76 \pm 0.14$ (statistical errors only).

The same procedure can be repeated after each relevant systematic check
listed in Table~\ref{tb:dspi0gam.syst} is performed.
The differences in $a_2$ values so obtained are added in quadrature to estimate
the total systematic uncertainty. The final result is:
\begin{equation}
a_2 = 0.76 \pm 0.14\;(\text{stat.}) \pm 0.06\;(\text{syst}) \;,
\end{equation}
where $a_2 = 0$ ($a_2 = 1$) corresponds to a $\DsFE$ of
helicity zero ($\pm 1$). This value of $a_2$ deviates from zero by 5.1
standard deviations, which strongly disfavors the 
$J^P = 0^-$ interpretation of the $\DsFE$ while remaining consistent
with a $J=1$ or higher interpretation of either parity

\section{\boldmath The $\Ds\piz\piz$ Final State}

\begin{figure}
\begin{center}
\includegraphics[width=0.49\linewidth]{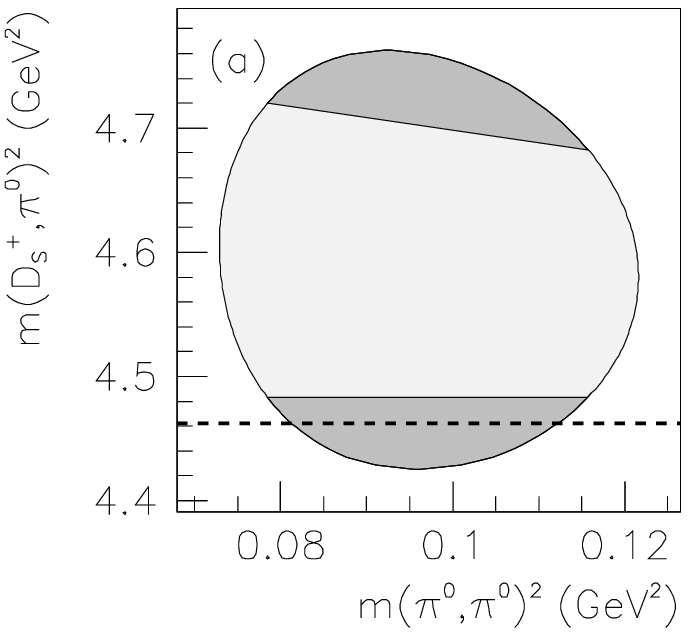}
\includegraphics[width=0.49\linewidth]{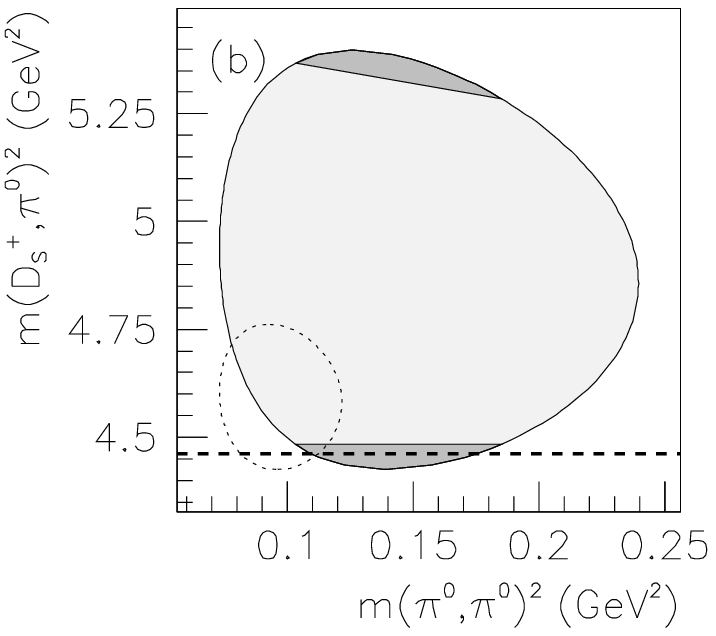}
\end{center}
\vskip -0.25in
\caption{\label{fg:ds2pi0.dalitzs}The Dalitz phase space available
to (a) the $\DsTT$ and (b) the $\DsFE$ mesons in the $\Ds\piz\piz$
final state. The dashed horizontal line corresponds to the $\DsTO$ mass.
The dark shaded regions are those parts of the phase
space removed by the requirement $m(\Ds\piz) > 2117.4$~\mevcc. 
The dotted curve in (b)
is the $\DsTT$ phase space drawn for comparison.
}
\end{figure}

The $\Ds\piz\piz$ final state contains potential contributions from
$\DsFE\to\DsTO\piz$ subresonant decay through the channel $\DsTO\to\Ds\piz$,
which has a branching ratio of 5.8\%~\cite{Eidelman:2004wy}. Since
this sub-resonant mode is more efficiently investigated using
the $\Ds\piz\gamma$ final state (as discussed in the previous section),
it is removed from the $\Ds\piz\piz$ sample by requiring the
$\Ds\piz$ invariant mass to be greater than 2117.4~\mevcc\  for both
$\piz$ candidates. This requirement excludes the
edges of the $\DsTT$ and $\DsFE$ phase spaces, as illustrated in
Fig.~\ref{fg:ds2pi0.dalitzs}. This figure also demonstrates how 
little phase space is available to the $\DsTT$ meson in this decay
in comparison to the $\DsFE$ meson.

\begin{figure}
\includegraphics[width=0.8\linewidth]{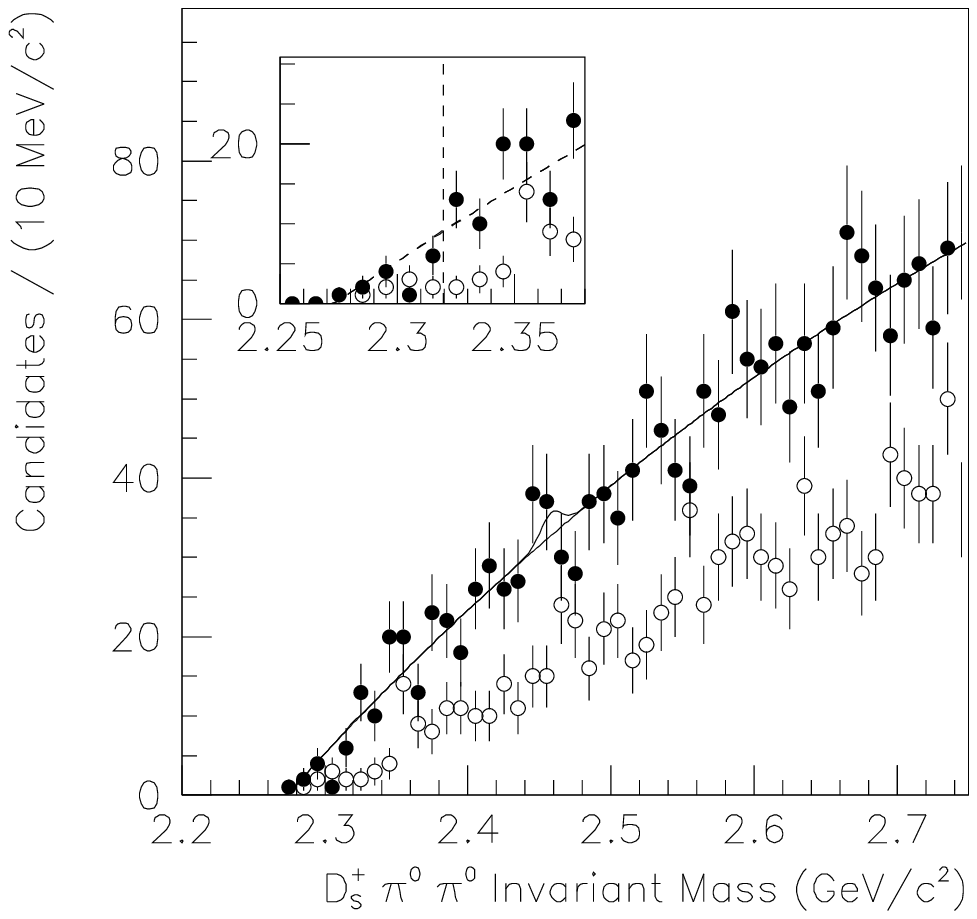}
\vskip -0.15in
\caption{\label{fg:ds2pi0.plotfitdatapaper}The invariant mass
distribution for (solid points) $\Ds\piz\piz$ candidates and (open points)
the equivalent using the $\Ds$ sidebands.
The curve represents the likelihood fit described in
the text. The insert focuses on the low mass region.
The dotted line in the insert indicates the $\DsTT$ mass.
}
\end{figure}

The invariant mass distribution of the selected $\Ds\piz\piz$
candidates is shown in Fig.~\ref{fg:ds2pi0.plotfitdatapaper}.
There is no evidence of $\DsTT$ or $\DsFE$ meson decay, nor is there
evidence of any structure in the background. The mass distribution is
fit using a likelihood function that consists of the smooth
background function $C(m)$ (Eq.~\ref{eq:cthres}) and 
$\DsTT$ and $\DsFE$ meson contributions, the latter two having
yields that are allowed to fluctuate to negative values. 
The shape of the signals is modeled by double Gaussians, 
the parameters of which
are determined by fits to MC samples. The masses of the $\DsTT$
and $\DsFE$ mesons are fixed to values measured in the previous
sections.

The result of the likelihood fit is shown in 
Fig.~\ref{fg:ds2pi0.plotfitdatapaper} and produces a $\DsTT$ ($\DsFE$)
raw yield of $0.2 \pm 3.9$ ($5 \pm 10$) (statistical errors only).
Efficiency corrections are calculated by applying a $\Ds\piz\piz$ 
candidate
weight that is inversely proportional to $p^*$-dependent selection
efficiency, calculated from MC samples. This procedure produces corrected
$\DsTT$ and $\DsFE$ meson yields of $1900 \pm 1600$ and $1300\pm 1200$, 
respectively (statistical errors only).

Various systematic uncertainties
are considered. According to MC simulation, the selection efficiency
varies by as much as 20\% across the Dalitz plot. Since no specific
Dalitz distribution is assumed, this variation is translated directly into
a multiplicative systematic uncertainty. An alternate
shape $C'_2(m)$ for the background is considered:
\begin{equation}
C'_2(m) = \sqrt{ m^2 - a_1^2 }\left( 1 + a_2 m + a_3m^3  \right) \;.
\end{equation}
A fit with this background increases the raw $\DsTT$  yield by 0.5
events. Detector resolution and $\DsTT$ and $\DsFE$ meson mass
variations have similar effects. The final results are
\begin{equation*}
\renewcommand{\arraystretch}{1.2}
\begin{array}{lcl@{}r@{\:}c@{\:}r@{\:}c@{\:}r@{}l}
\bar{\sigma}(\DsTT\to\Ds\piz\piz) 
&=& (&8.7 &\pm&   6.9  &\pm&   5.0&)
\;\text{fb}  \\
\bar{\sigma}(\DsFE\to\Ds\piz\piz) 
&=& (&5.5 &\pm&  5.4  &\pm&  2.4&)
\;\text{fb},
\end{array}
\end{equation*}
where the first error is statistical and the second is systematic.

\section{\boldmath The $\Ds\gamma\gamma$ Final State}

It is assumed that there are three possible
contributions to the decay of the $\DsTT$
and $\DsFE$ mesons to the $\Ds\gamma\gamma$ final state:
\begin{itemize}
\item
Two body decay to $\Ds\piz$ followed by $\piz\to\gamma\gamma$.
\item
Subresonant decay to $\DsTO\gamma$ followed by $\DsTO\to\Ds\gamma$.
\item
Non-resonant decay directly to $\Ds\gamma\gamma$.
\end{itemize}
The $\Ds\piz$ final state, already studied in Section~\ref{sec:dspi0},
is removed from the $\Ds\gamma\gamma$ sample by the $\gamma$
selection requirements. Potential
background from $\piz$ mass tails is removed by further requiring
the $\gamma\gamma$ invariant mass to be less than 100~\mevcc\  or
greater than 170~\mevcc.
The remaining two $\Ds\gamma\gamma$ contributions are treated by dividing the
sample into two portions, one rich in $\DsTO$ decay (referred to as the
$\DsTO$ sample),
and the remainder (referred to as the non-resonant sample). 
A $\Ds\gamma\gamma$ candidate is placed in the
$\DsTO$ sample if either $\gamma$ produces a $\Ds\gamma$ invariant mass within
15~\mevcc\  of the PDG value for the $\DsTO$ mass 
(2112.4~\mevcc)~\cite{Eidelman:2004wy}.

\begin{figure}
\includegraphics[width=\linewidth]{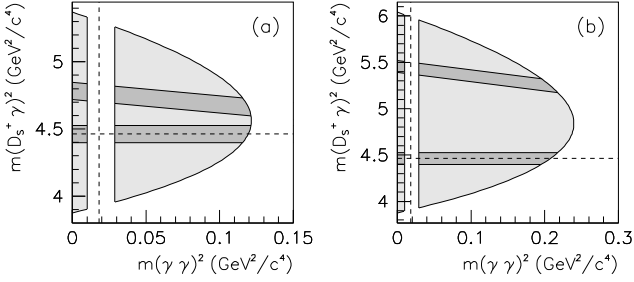}
\vskip -0.15in
\caption{\label{fg:ds2gam.dalitzs}The Dalitz phase space corresponding to
the $\Ds\gamma\gamma$ samples for (a) $\DsTT$ and (b) $\DsFE$ mesons.
The dark shaded and light shaded regions correspond to the $\DsTO$
and non-resonant samples, respectively. The horizontal (vertical) dashed line
indicates the $\DsTO$ ($\piz$) mass.
}
\end{figure}

\begin{figure}
\includegraphics[width=\linewidth]{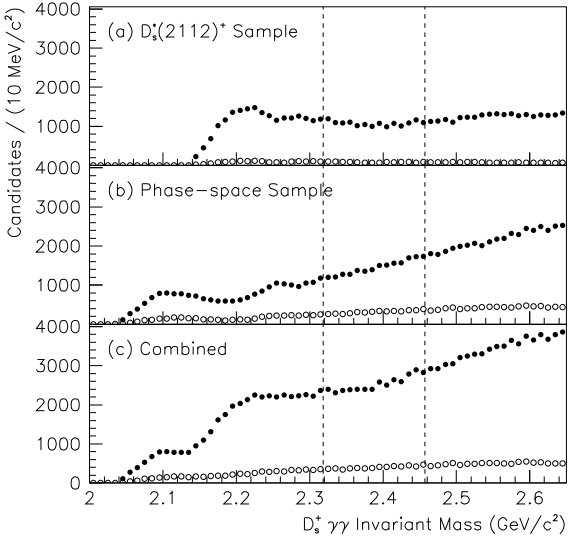}
\vskip -0.15in
\caption{\label{fg:ds2gam.datamanypaper}The $\Ds\gamma\gamma$
invariant mass distribution (solid points)
for candidates in (a) the $\DsTO$ sample, (b)
the non-resonant sample, and (c) either sample. The open points are
the corresponding distributions for the $\Ds$ sideband.
The vertical dashed lines indicate the $\DsTT$ and $\DsFE$ meson
masses.
}
\end{figure}

Figure~\ref{fg:ds2gam.dalitzs} illustrates how the 
the $\Ds\gamma\gamma$ candidates are divided  in terms of the 
phase space of $\DsTT$ and $\DsFE$ meson decay.
The $\Ds\gamma\gamma$ invariant mass distribution of the two samples
is shown in Fig.~\ref{fg:ds2gam.datamanypaper}. No clear $\DsTT$ or
$\DsFE$ signal is observed. The mass distributions contain structure
associated both with the $\piz$ veto requirements and with the
$\DsTO$ sample selection requirements. Much of this structure can be
avoided by applying likelihood fits in a restricted mass range.
Two types of background that produce structure
that cannot be so avoided are described below.

\begin{figure}
\includegraphics[width=0.49\linewidth]{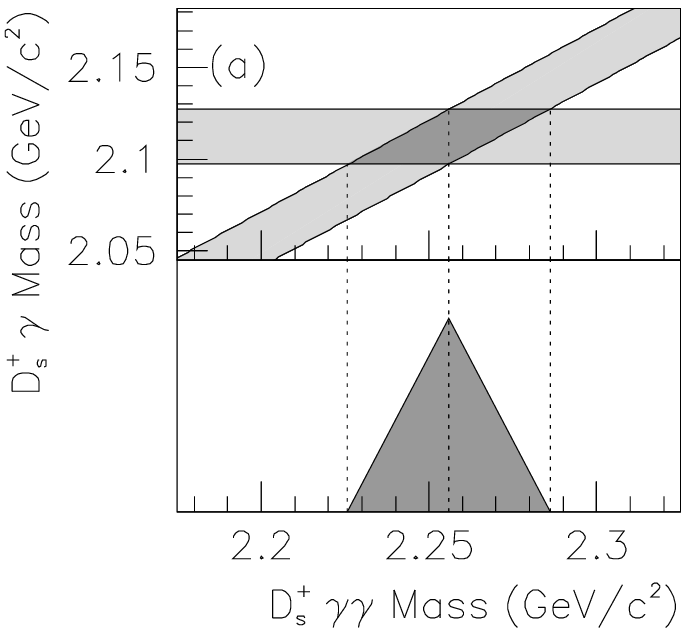}
\includegraphics[width=0.49\linewidth]{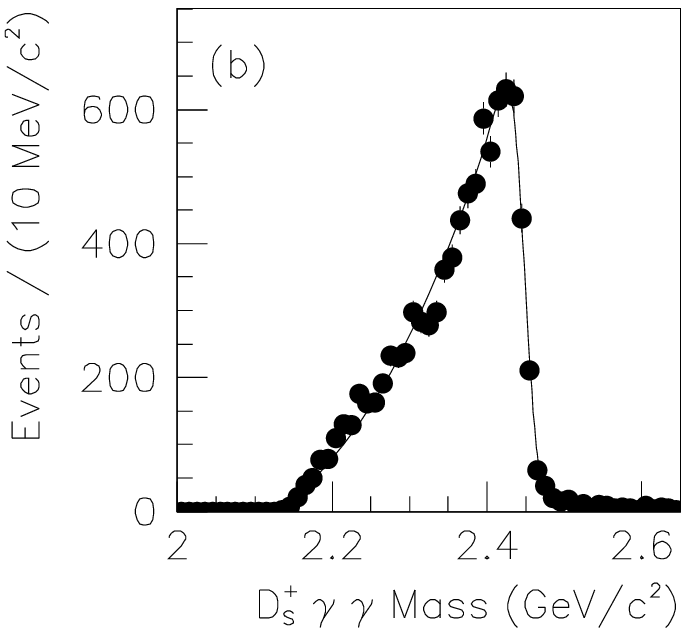}
\vskip -0.15in
\caption{\label{fg:ds2gam.backs}(a) Shown on top as a function
of $\Ds\gamma$ and $\Ds\gamma\gamma$ masses are the bands corresponding
to the $\DsTO$ sample requirement ($\Ds\gamma$ mass within 15~\mevcc\  of
the $\DsTO$ mass) and the effect of this requirement on the other $\gamma$
candidate.
The overlap of the two bands produces the shape shown below. (b) The
$\Ds\gamma\gamma$ mass distribution of the $\DsFE\to\DsTO\piz$ reflection
from a MC sample. The curve is the fit described in the text.
}
\end{figure}

Either of the two $\gamma$ candidates in each 
$\Ds\gamma\gamma$ combination provides two opportunities for 
the $\Ds\gamma\gamma$
combination to be placed into the $\DsTO$ sample. At a specific
value of $\Ds\gamma\gamma$ invariant mass, however, if one $\gamma$
candidate falls inside the $\DsTO$ mass window, the other $\gamma$ 
is likely to do the same, due to kinematics. This produces a
deficit of candidates at this mass in the $\DsTO$ sample along with a 
corresponding excess of candidates in the non-resonant sample.
The $\Ds\gamma\gamma$ invariant mass distribution of this excess 
can be approximated by the triangular shape centered at 2255.9~\mevcc 
shown in Fig.~\ref{fg:ds2gam.backs}(a).

The second background source, which only appears in the $\DsTO$ sample,
is a reflection from $\DsFE\to\DsTO\piz$ decay in which one $\gamma$
particle from the $\piz$ decay is ignored. This reflection is
suppressed but not eliminated
by the requirement that each $\gamma$ candidate not belong to
a fiducial $\piz$ candidate. The MC prediction of the shape of this reflection
is shown in Fig.~\ref{fg:ds2gam.backs}(b). The shape is accurately
modeled by a quadratic polynomial bounded on two sides
and smeared by a double Gaussian.

The unbinned maximum-likelihood
fit to the $\Ds\gamma\gamma$ mass distribution of the $\DsTO$ sample
is performed between masses of 2.22 and 2.70~\gevcc.
This mass distribution has a cusp near a mass of 2.4~\gevcc\  that is
difficult to describe using a simple polynomial. Instead, 
the combinatorial background is parameterized using an empirical
function $K(m)$
composed of a line and a parabola that intersect at one point.
To make the function smooth, a cubic spline $K_S(m)$ 
is used near the intersection point ($m=a_1$) to extrapolate
between these two polynomials:
\begin{equation}
K(m) = \left\{ 
\begin{array}{ll}
a_2 + a_3 m           & m < a_1 - \delta \\
K_S(m)                & \left| m - a_1 \right| < \delta \\
a_2 + a_4 m + a_5 m^2 & m > a_1 + \delta
\end{array}
\right.\;,
\label{eq:ds2gam.cusp}
\end{equation}
where the value $\delta = 20$~\mevcc\  is chosen to 
approximately match the resolution.

The reconstructed mass distributions for the hypothetical decay of $\DsTT$ and
$\DsFE$ mesons to $\Ds\gamma\gamma$ and $\DsTO\gamma$ 
are modeled by the functional
form of Eq~\ref{eq:modlorentz2}, the parameters of which are determined
by fits to the corresponding MC samples. 

\begin{figure}
\begin{center}
\includegraphics[width=\linewidth]{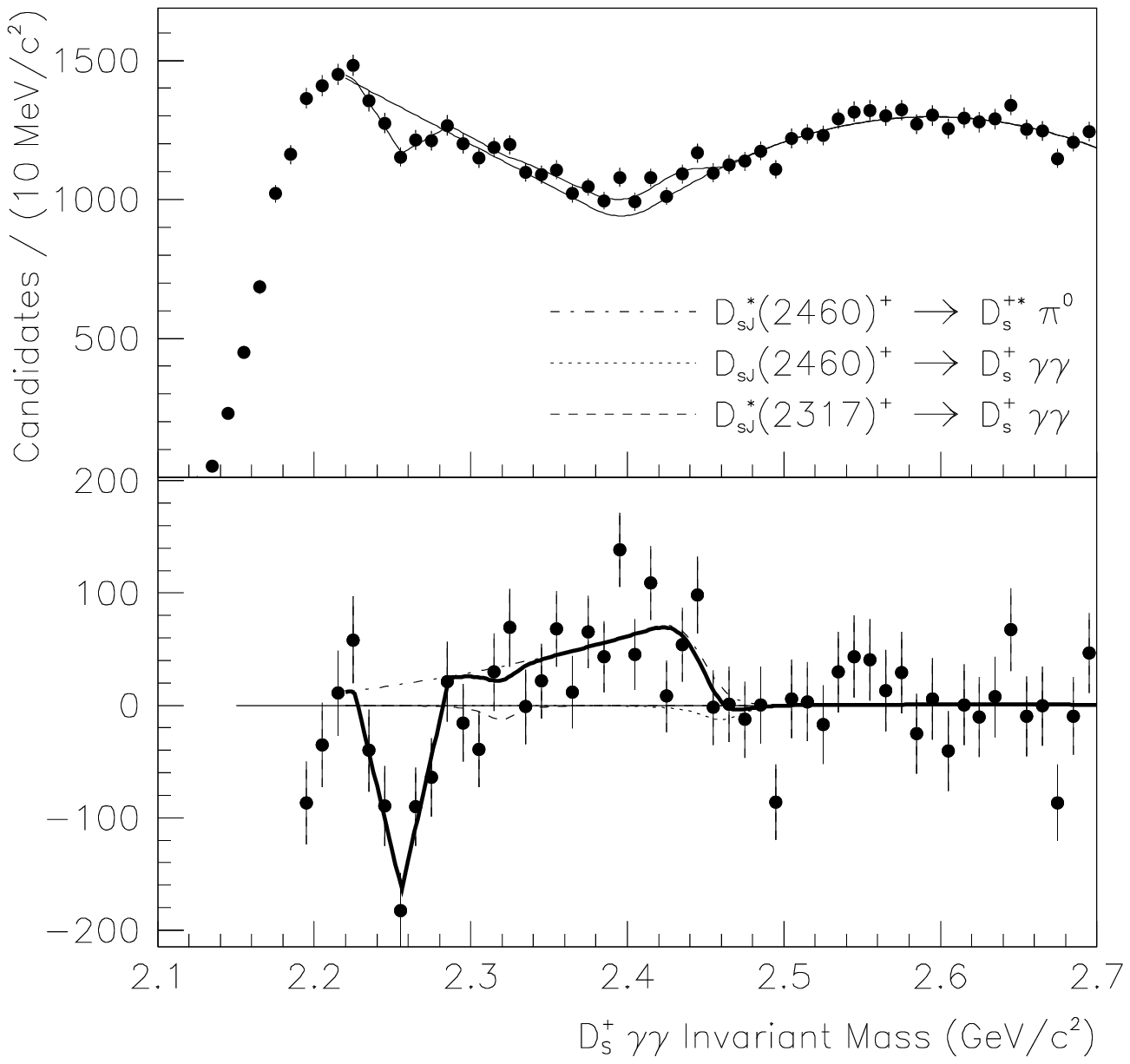}
\end{center}
\vskip -0.25in
\caption{\label{fg:ds2gam.plotfitdatapaper2}The
$\Ds\gamma\gamma$ invariant mass distribution for the $\DsTO$ sample
is shown on top.
The thick curve is the result of the likelihood fit. 
The mass distribution after subtracting the portion assigned by the
fit to combinatorial background is shown on bottom.
}
\end{figure}

The unbinned likelihood fit to the $\Ds\gamma\gamma$ mass distribution
of the $\DsTO$ sample includes five components:
\begin{enumerate}
\item
$\DsTT\to\DsTO\gamma$ decay (hypothetical). 
\item
$\DsFE\to\DsTO\gamma$ decay (hypothetical). 
\item
A deficit of events of the shape described in Fig.~\ref{fg:ds2gam.backs}(a)
but of variable size.
\item
A contribution from the $\DsFE\to\DsTO\piz$ reflection, of fixed shape
based on the fit of Fig.~\ref{fg:ds2gam.backs}(b) and with a
yield consistent with the results of Section~\ref{sec:dspi0gam}.
\item
A background described by the function of Eq.~\ref{eq:ds2gam.cusp}.
\end{enumerate}
The result of applying the fit function on
the $\DsTO$ sample is shown in Fig.~\ref{fg:ds2gam.plotfitdatapaper2}.
A raw $\DsTT$ ($\DsFE$) yield of $-40 \pm 110$ ($-50 \pm 140$) is obtained
(statistical errors only).

The proximity of the $\DsFE\to\DsTO\piz$ reflection to potential
$\DsTT$ and $\DsFE$ signals coupled with the unknown
shape of the combinatorial background
leads to large uncertainties in the fit results. For example, if a
third-order polynomial is used in place of the function $K(m)$ to
describe the smooth background, the $\DsTT$ and $\DsFE$ raw yields 
increase by 260 and 150 candidates, respectively. Although the result of this
fit is a less faithful representation of the mass distribution of the
data, as a conservative estimate
the entire difference is quoted as a systematic uncertainty.
Other systematic checks performed include a variation of the range in
$\Ds\gamma\gamma$
mass over which the fit is applied, consideration of uncertainties in detector
resolution, and variations of the $\DsTT$ and $\DsFE$ masses and the
relative size of the $\DsFE\to\DsTO\piz$ reflection. 

The raw $\DsTT$ and $\DsFE$ yields
are corrected for selection efficiency by weighting each $\Ds\gamma\gamma$
combination by the inverse of the $p^*$-dependent selection efficiency,
determined using a MC sample. The results for the $\DsTO\gamma$
final state are:
\begin{equation*}
\renewcommand{\arraystretch}{1.2}
\begin{array}{lcl@{}r@{\:}c@{\:}r@{\:}c@{\:}r@{}l}
\bar{\sigma}(\DsTT\to\DsTO\gamma) 
&=& (&-0.5 &\pm& 3.2  &\pm&  8.1&)
\;\text{fb}  \\
\bar{\sigma}(\DsFE\to\DsTO\gamma) 
&=& (&-0.9 &\pm& 3.5  &\pm&  4.1&)
\;\text{fb},
\end{array}
\end{equation*}
where the first error is statistical and the second is systematic.

\begin{figure}
\begin{center}
\includegraphics[width=\linewidth]{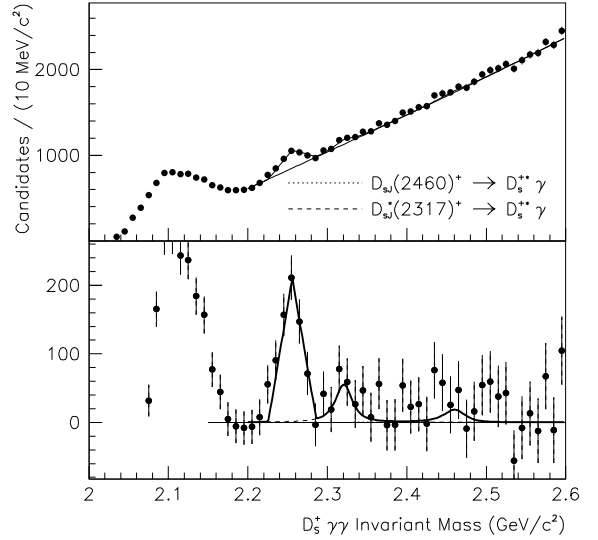}
\end{center}
\vskip -0.25in
\caption{\label{fg:ds2gam.plotfitdatapaper1}The
$\Ds\gamma\gamma$ invariant mass distribution for the non-resonant sample
is shown on top.
The thick curve is the result of the likelihood fit. 
The mass distribution after subtracting the portion assigned by the
fit to combinatorial background is shown on bottom.
}
\end{figure}

The $\Ds\gamma\gamma$ mass distribution of the phase-space subsample
has a cusp that is more pronounced than in the $\DsTO\gamma$ sample
but at a mass well below the $\DsTT$ mass. This cusp
is avoided by restricting the fit to masses above
2.2~\gevcc. In this mass range a simple polynomial is sufficient to describe
combinatorial background.
The unbinned likelihood fit includes four components:
\begin{enumerate}
\item
$\DsTT\to\Ds\gamma\gamma$ decay (hypothetical). 
\item
$\DsFE\to\Ds\gamma\gamma$ decay (hypothetical). 
\item
An excess of events of the shape described in Fig.~\ref{fg:ds2gam.backs}(a)
but of variable size.
\item
Combinatorial background represented by a third-order polynomial.
\end{enumerate}
The result of applying the fit function on
the non-resonant sample is shown in Fig.~\ref{fg:ds2gam.plotfitdatapaper1}.
A raw $\DsTT$ ($\DsFE$) yield of $190 \pm 120$ ($80 \pm 160$) is obtained
(statistical errors only).

As in the $\DsTO$ sample, raw $\DsTT$ and $\DsFE$ yields
are corrected for selection efficiency by weighting each $\Ds\gamma\gamma$
combination by the inverse of the $p^*$-dependent selection efficiency.
This efficiency was determined using a MC sample that simulated
the non-resonant decays of the $\DsTT$ and $\DsFE$ mesons to
$\Ds\gamma\gamma$ using phase space (such that the Dalitz plot was
evenly populated). To simulate an alternate decay model, the MC samples
were weighted to form a $\cos^2\vartheta_{D\gamma}$ distribution,
where $\vartheta_{D\gamma}$ is the angle between the $\Ds$ and
each $\gamma$ in the $\DsTT$ and $\DsFE$ center-of-mass frame. This weighting
had the effect of reducing the selection efficiency by 25\% and 15\%
for the $\DsTT$ and $\DsFE$ mesons, respectively.

Other systematic checks include variations in detector resolution,
the $\DsTT$ and $\DsFE$ meson masses, and the fit range.
The final results for the non-resonant $\Ds\gamma\gamma$ final state are:
\begin{equation*}
\renewcommand{\arraystretch}{1.2}
\begin{array}{lcl@{}r@{\:}c@{\:}r@{\:}c@{\:}r@{}l}
\bar{\sigma}(\DsTT\to\Ds\gamma\gamma) 
&=& (&7.4 &\pm& 4.5  &\pm&  2.2&)
\;\text{fb}  \\
\bar{\sigma}(\DsFE\to\Ds\gamma\gamma) 
&=& (&3.5 &\pm& 4.3  &\pm&  1.7&)
\;\text{fb},
\end{array}
\end{equation*}
where the first error is statistical and the second is systematic.

\section{\boldmath The $\Ds\pipm$ Final States}

\begin{figure}
\begin{center}
\includegraphics[width=\linewidth]{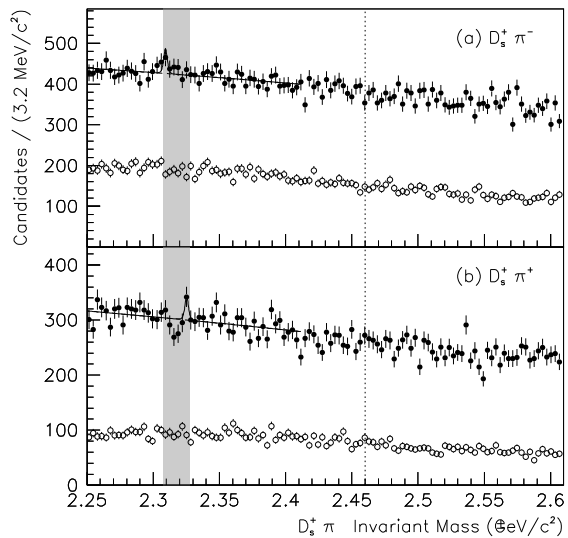}
\end{center}
\vskip -0.25in
\caption{\label{fg:dspi.fitpaper}The invariant mass distributions
(solids points)
of (a) $\Ds\pim$ and (b) $\Ds\pip$ candidates
and (open points) the equivalent for the $\Ds$ sidebands.
The shaded regions indicate the range of assumed
$\DsTTdc$ and $\DsTTz$ masses. The vertical dotted line marks
the $\DsFE$ mass. The curves are 
the fits described
in the text that produce the largest yield within the shaded region.
}
\end{figure}

The invariant mass distributions of the $\Ds\pim$ and $\Ds\pip$
candidates are shown in Fig.~\ref{fg:dspi.fitpaper} for masses between
2.25 and 2.61~\gevcc. No structure is apparent and both distributions
can be adequately described by a second-order polynomial. Since no signal
is apparent, the remaining task is to place limits on the yield
of hypothetical
doubly-charged ($\DsTTdc$) or neutral ($\DsTTz$) partners of the
$\DsTT$ meson. To do so requires a few assumptions:
\begin{itemize}
\item
The intrinsic width $\Gamma$ of either the $\DsTTz$ or $\DsTTdc$ particle
is too small to be resolved by the detector.
\item
The mass of the $\DsTTz$ or $\DsTTdc$ particle is within $\pm 10$~\mevcc\  of
the $\DsTT$ mass.
\end{itemize}

A series of unbinned likelihood fits based on these assumptions
are applied to the $\Ds\pim$ and 
$\Ds\pip$ mass distributions. Included in these fits is a $\DsTTz$ or
$\DsTTdc$ signal modeled using a line shape extracted 
from a fit to a $\DsTTz\to\Ds\pim$ MC sample. According to this fit,
the mass resolution is approximately 1.3~\mevcc. 
To avoid potential statistical biases, the mass of each
hypothetical particle is fixed at a specific value for each fit.
Several such fits are applied with the assumed mass
placed between 2307.3 and 2327.3~\mevcc\
at intervals of 1~\mevcc . The fits that produce the largest
yields are shown in Fig.~\ref{fg:dspi.fitpaper}. 

The two final states discussed in this section involve only charged
particles and are thus subject to relatively small systematic
uncertainties. The largest uncertainty arises from the assumed
shape of the background and is estimated by substituting 
a third-order polynomial for the second-order one. 
There is also a 1.3\% relative
uncertainty in the reconstruction efficiency of each $\pipm$ candidate.
The results from the fits that produce the largest yields are:
\begin{equation*}
\renewcommand{\arraystretch}{1.2}
\begin{array}{lcl@{}r@{\:}c@{\:}r@{\:}c@{\:}r@{}l}
\bar{\sigma}(\DsTTz\to\Ds\pim) 
&=& (&1.07 &\pm& 0.44  &\pm&  0.10&)
\;\text{fb} \\
\bar{\sigma}(\DsTTdc\to\Ds\pip) 
&=& (&0.74 &\pm& 0.37  &\pm&  0.07&)
\;\text{fb},
\end{array}
\end{equation*}
where the first error is statistical and the second is systematic.
These results will be used to calculate upper limits on the 
cross section.

\section{\boldmath The $\Ds\pip\pim$ Final State}
\label{sec:dspipm}

\begin{figure}
\begin{center}
\includegraphics[width=\linewidth]{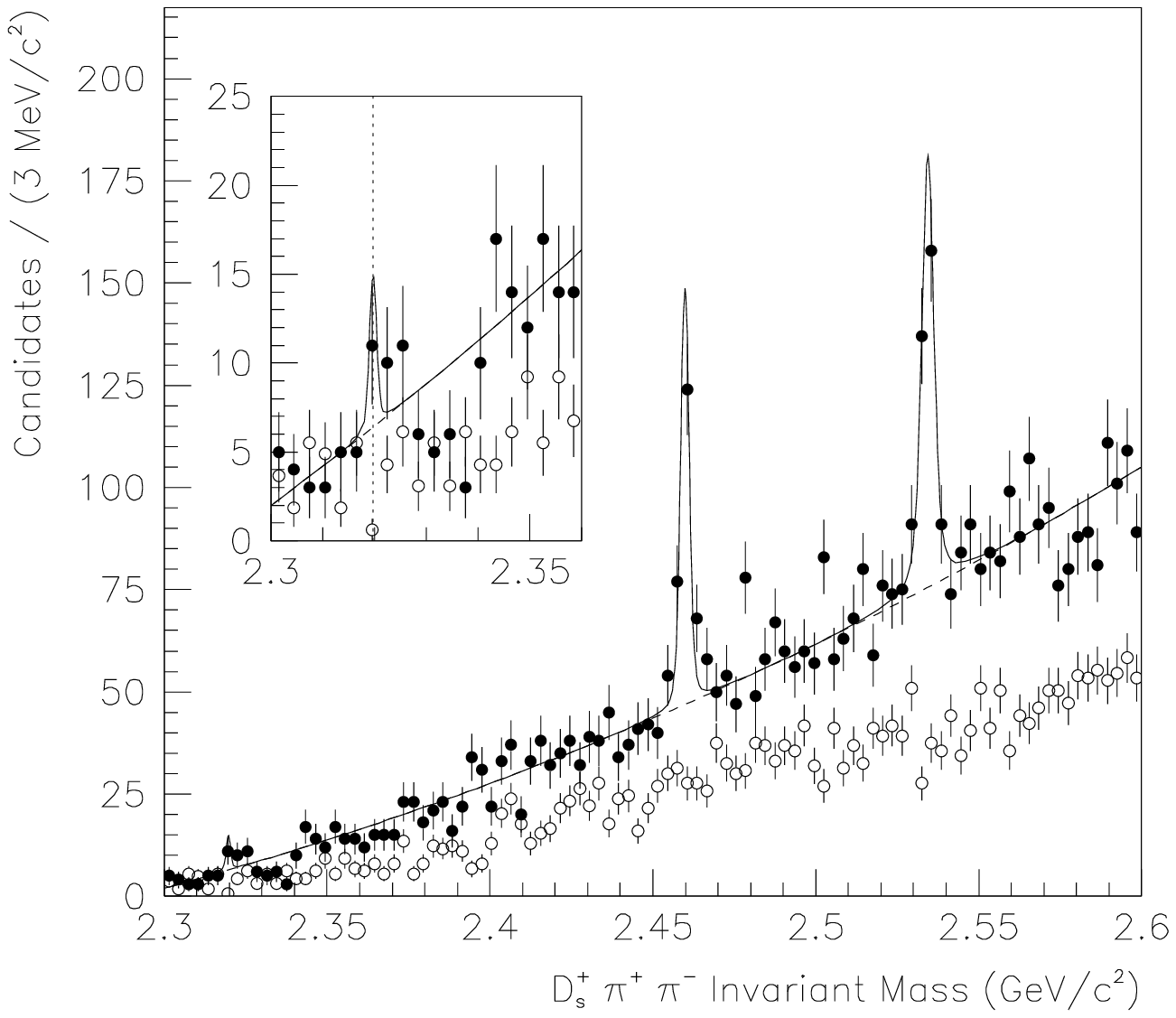}
\end{center}
\vskip -0.25in
\caption{\label{fg:dspipm.fitpaper}The invariant mass distribution
of (solid points) $\Ds\pip\pim$ candidates
and (open points) the equivalent using the $\Ds$ sidebands. 
The curve is the fit described in the
text. The insert focuses on the low mass region.
The dotted line in the insert indicates the $\DsTT$ mass.
}
\end{figure}

\begin{figure}
\begin{center}
\includegraphics[width=0.49\linewidth]{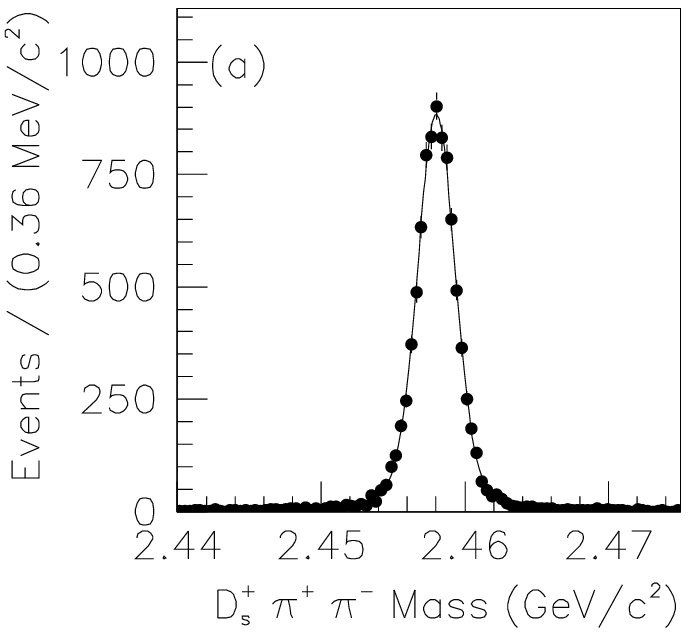}
\includegraphics[width=0.49\linewidth]{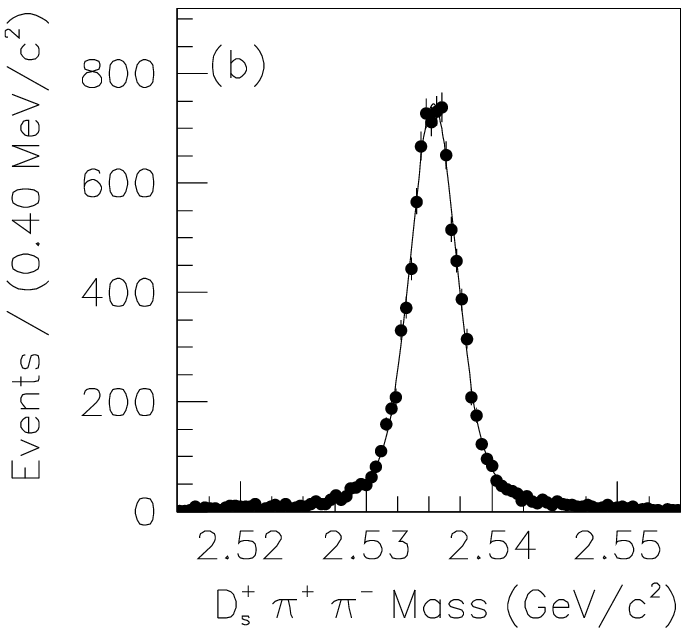}
\end{center}
\vskip -0.25in
\caption{\label{fg:dspipm.signals}The invariant mass distributions
of MC samples of (a) $\DsFE$ and (b) $\DsTS$ decay to $\Ds\pip\pim$.
The curves are the functions described in the text.
}
\end{figure}

The invariant mass distribution of the $\Ds\pip\pim$ candidates is
shown in Fig.~\ref{fg:dspipm.fitpaper}. Clear peaks from $\DsFE$ 
and $\DsTS$ decay are apparent. The distribution has no additional structure
of any statistical significance. To determine the mass and yield of
the $\DsFE$ meson, and to place limits on $\DsTT$ decay, an unbinned
likelihood fit is applied to the $\Ds\pip\pim$ mass distribution.
This fit includes the following contributions:
\begin{itemize}
\item
$\DsTT\to\Ds\pip\pim$ decay (hypothetical).
\item
$\DsFE\to\Ds\pip\pim$ decay.
\item
$\DsTS\to\Ds\pip\pim$ decay.
\item
A third-order polynomial to describe the background.
\end{itemize}
Each signal decay mode is described by a PDF consisting of three
Gaussians with a common mean, the parameters of which are determined using
fits to MC samples (generated with $\Gamma = 0.1$~\mev). 
The fits for the $\DsFE$ and $\DsTS$ mesons
are shown in Fig.~\ref{fg:dspipm.signals} and correspond to
mass resolutions of approximately 1.3 and 1.9~\mevcc, respectively.
The $\DsFE$ and $\DsTS$ masses are allowed to vary in the fit
and the $\DsTT$ mass is fixed to the value determined
in Section~\ref{sec:dspi0}.

The result of the likelihood fit is shown in Fig.~\ref{fg:dspipm.fitpaper}.
A $\DsFE$ mass of $(2459.7 \pm 0.2)$~\mevcc\ and 
$\DsTS$ mass of $(2534.3 \pm 0.3)$~\mevcc\
are obtained from raw yields of $123\pm 15$ and $193\pm 22$,
respectively (statistical errors only). A raw $\DsTT$ yield
of $6 \pm 3$ is also obtained, consistent with zero.

The uncertainties in the $\DsFE$ and $\DsTS$ masses are summarized
in Table~\ref{tb:dspipm.syst}. The largest uncertainty (0.6~\mevcc)
is in the assumed $\Ds$ mass. In comparison, uncertainties associated
with the likelihood fit, such as the background shape or mass
resolution, are relatively small. The remaining uncertainties are
attributed to potential momentum biases in the tracking detectors.

\begin{table}
\caption{\label{tb:dspipm.syst}A summary of systematic uncertainties
for the $\DsFE$ and $\DsTS$ masses from the analysis of the
$\Ds\pip\pim$ final state.
}
\begin{center}

\begin{ruledtabular}
\begin{tabular}{lrr}
    & \multicolumn{2}{c}{Mass Uncertainty (\mevcc)} \\
\cline{2-3}
Source & $\DsFE$ & $\DsTS$ \\
\hline
$\Ds$ mass              & 0.60 & 0.60 \\
Signal line shape       & 0.03 & 0.01 \\
Background function     & 0.01 & 0.01 \\
Solenoid field          & 0.05 & 0.08 \\
Magnetization           & 0.02 & $<0.01$ \\
Energy-loss correction  & 0.52 & 0.30 \\
$\phi$ dependence       & 0.01 & 0.01 \\
\hline
Quadrature Sum & 0.80 & 0.68 \\
\end{tabular}
\end{ruledtabular}

\end{center}
\end{table}

Uncertainties in tracking have several sources, discussed
in detail as part of a recent measurement of the $\Lambda_c$ mass
from this collaboration~\cite{Aubert:2005gt}. A similar treatment
of these uncertainties is reproduced for this analysis. The magnetic
field of the \babar\  detector is known to high precision,
and variations of both the overall strength of the solenoidal field
and of magnetization effects of \pep2\  beam elements produce
no significant variation in measured $\Ds\pip\pim$ mass.
Residual $\phi$-dependent momentum tracking biases are also insignificant.
The largest uncertainty arises from material inside the tracking
volume and the effect of this material on energy loss corrections.
Studies of large samples of $\Lambda$ and $K_S$ decays suggest that
either the amount of material and/or atomic weight composition of
the SVT is underestimated by approximately 20\%. MC studies are used 
to estimate the bias introduced by this
underestimation. A correction of $0.46$~\mevcc\   and $0.29$~\mevcc\
for the $\DsFE$ and $\DsTS$ masses is calculated based on these
studies. The final results are:
\begin{equation*}
\renewcommand{\arraystretch}{1.2}
\begin{array}{lcl@{}r@{\:}c@{\:}r@{\:}c@{\:}r@{}l}
m(\DsFE)
&=& (&2460.2 &\pm& 0.2  &\pm&  0.8&)
\;\text{\mevcc}  \\
m(\DsTS)
&=& (&2534.6 &\pm& 0.3  &\pm&  0.7&)
\;\text{\mevcc},
\end{array}
\end{equation*}
where the first error is statistical and the second is systematic.

The $\DsFE$ and $\DsTS$ signal PDF used in the likelihood fit of 
Fig.~\ref{fg:dspi0gam.plotfitpaper} was based on intrinsic widths
of $\Gamma = 0.1$~\mev. 
After applying the same likelihood-integration technique
described in Section~\ref{sec:dspi0} for the $\Ds\piz$ final state,
and after considering systematic uncertainties in the background shape
and mass reconstruction resolution,
the result is a 95\% CL limit of $\Gamma<3.5$~\mev\ and $\Gamma<2.5$~\mev\ 
for the $\DsFE$ and $\DsTS$ mesons.

To estimate a total $\DsTT$ yield, a weight inversely proportional to the
$p^*$-dependent $\DsTT$ selection efficiency is applied to
each $\Ds\pip\pim$ combination and the likelihood fit repeated.
A similar process is used for the $\DsFE$ and $\DsTS$ yields.
The results are yields of
$60 \pm 40$, $760\pm 110$, and $1210\pm 160$
for the $\DsTT$, $\DsFE$, and $\DsTS$ mesons, respectively, and
for $p^*>3.2$~\gevc\ (statistical errors only).

The systematic uncertainties in the $\DsFE$ and $\DsTS$ yields are
summarized in Table~\ref{tb:dspipm.syst2}. The cross section results are:
\begin{equation*}
\renewcommand{\arraystretch}{1.2}
\begin{array}{lcl@{}r@{\:}c@{\:}r@{\:}c@{\:}r@{}l}
\bar{\sigma}(\DsTT\to\Ds\pip\pim) 
&=& (&0.3 &\pm& 0.2  &\pm&  0.0&)
\;\text{fb}  \\
\bar{\sigma}(\DsFE\to\Ds\pip\pim) 
&=& (&3.3 &\pm& 0.5  &\pm&  0.3&)
\;\text{fb}, \\
\bar{\sigma}(\DsTS\to\Ds\pip\pim) 
&=& (&5.2 &\pm& 0.7  &\pm&  0.4&)
\;\text{fb},
\end{array}
\end{equation*}
where the first error is statistical and the second is systematic.

\begin{table}
\caption{\label{tb:dspipm.syst2}A summary of systematic uncertainties
for the $\DsFE$ and $\DsTS$ yields from the analysis of the
$\Ds\pip\pim$ final state.
}
\begin{center}

\begin{ruledtabular}
\begin{tabular}{lrr}
    & \multicolumn{2}{c}{Relative Yield (\%)} \\
\cline{2-3}
Source & $\DsFE$ & $\DsTS$ \\
\hline
Tracking Efficiency              & 2.7 & 2.7 \\
Signal line shape                & 2.7 & 0.3 \\
Background function              & 6.1 & 5.6 \\
$\Ds$ efficiency                 & 3.9 & 3.9  \\
$\Ds\to\phi\pip$ yield           & 3.9 & 3.9  \\
\hline
Quadrature Sum & 9.1 & 8.3 \\
\end{tabular}
\end{ruledtabular}

\end{center}
\end{table}

\section{Combined Mass and Width Results}

The mass and width results for the $\DsTT$, $\DsFE$, and
$\DsTS$ mesons are summarized in Table~\ref{tb:masses}. The $\DsTT$
and $\DsTS$ mesons are observed in only one decay mode covered by this
analysis; those portions of this table are copied unchanged
from the respective sections of this paper. The $\DsFE$ mass
is the average of that obtained from the $\Ds\gamma$, $\Ds\piz\gamma$,
and $\Ds\pip\pim$ final states, although the latter measurement
dominates in the average due to superior systematic uncertainties.
This average is calculated using a $\chi^2$ method that
properly accounts for correlations among the systematic
uncertainties in the three measurements.
The limit on the intrinsic $\DsFE$ width $\Gamma$ is taken as the 
best limit obtained from these three decay modes.

\begin{table}
\caption{\label{tb:masses}A summary of the combined
mass and width results. For the masses, the first quoted uncertainty
is statistical and the second is systematic. The limits on 
the intrinsic width $\Gamma$ are at 95\% CL.
}
\begin{center}

\begin{ruledtabular}
\begin{tabular}{l@{\hspace{0.5cm}}r@{$\:\pm\:$}r@{$\:\pm\:$}r@{\hspace{0.5cm}}r}
Particle & \multicolumn{3}{c}{Mass (\mevcc)} & $\Gamma$ (\mev) \\
\hline
$\DsTT$ & 2319.6 & 0.2 & 1.4 & $<3.8$ \\
$\DsFE$ & 2460.1 & 0.2 & 0.8 & $<3.5$ \\
$\DsTS$ & 2534.6 & 0.3 & 0.7 & $<2.5$ \\
\end{tabular}
\end{ruledtabular}

\end{center}
\end{table}

\section{Yields and Branching Ratios}

The eighteen decay 
yields $\bar{\sigma}$ measured in this paper are collected in 
Table~\ref{tb:yields}. As described by Eq.~\ref{eq:sigmabar}, 
these numbers correspond to total
yields for the decay of mesons having a center-of-mass momentum
$p^*$ of at least 3.2~\gevc\  to a final state that includes a
$\Ds$ meson that decayed to $\phi\pip$.

\begin{table}
\caption{\label{tb:yields}A summary of yield results.
All cross sections are calculated for $p^* > 3.2$~\gevc. 
The first quoted uncertainty for the central value
is statistical and the second is systematic. The limits
correspond to 95\% CL.}
\begin{ruledtabular}
\renewcommand{\baselinestretch}{1.3}
\begin{tabular}{llr@{$\:\pm\:$}r@{$\:\pm\:$}rr}
\multicolumn{2}{l}{Decay Mode} 
     & \multicolumn{3}{c}{Central Value (fb)} & Limit (fb)  \\
\hline
\multicolumn{6}{l}{$\sigma(\DsTT)\mathcal{B}(\DsTT\to X) \mathcal{B}( \Ds\to\phi\pip )$} \\
\hspace{5pt}
& $\Ds\piz$          & $115.8$ & $  2.9$ & $  8.7$ &    ---   \\
& $\Ds\gamma$        & $ -2.4$ & $  2.3$ & $  8.9$ & $< 15.7$ \\
& $\Ds\pi^0\pi^0$    & $  8.7$ & $  6.9$ & $  5.0$ & $< 29.0$ \\
& $\Ds\gamma\gamma$  & $  7.4$ & $  4.5$ & $  2.2$ & $< 20.6$ \\
& $\DsTO\gamma$      & $ -0.5$ & $  3.2$ & $  8.1$ & $< 16.7$ \\
& $\Ds\pi^+\pi^-$    & $  0.3$ & $  0.2$ & $  0.0$ & $<  0.6$ \\
\multicolumn{6}{l}{$\sigma(\DsFE)\mathcal{B}(\DsFE\to X) \mathcal{B}( \Ds\to\phi\pip )$} \\
& $\Ds\piz$          & $ -1.0$ & $  1.4$ & $  0.1$ & $<  1.7$ \\
& $\Ds\gamma$        & $ 14.4$ & $  1.0$ & $  1.4$ &    ---   \\
& $\Ds\piz\gamma$ [a]& $ 42.7$ & $  3.5$ & $  4.2$ &    ---   \\
& $\DsTO\piz$        & $ 41.6$ & $  5.1$ & $  5.0$ &    ---   \\
& $\DsTT\gamma$      & $  1.1$ & $  5.1$ & $  5.0$ & $< 15.2$ \\
& $\Ds\pi^0\pi^0$    & $  5.5$ & $  5.4$ & $  2.4$ & $< 28.5$ \\
& $\Ds\gamma\gamma$  & $  3.5$ & $  4.3$ & $  1.7$ & $< 13.2$ \\
& $\DsTO\gamma$      & $ -0.9$ & $  3.5$ & $  4.1$ & $<  9.7$ \\
& $\Ds\pi^+\pi^-$    & $  3.3$ & $  0.5$ & $  0.3$ &    ---   \\
\hline
\multicolumn{6}{l}{$\sigma(\DsTTdc)\mathcal{B}(\DsTTdc\to X) \mathcal{B}( \Ds\to\phi\pip )$} \\
& $\Ds\pi^+$         & \multicolumn{3}{c}{---}     & $<  1.5$ \\
\hline
\multicolumn{6}{l}{$\sigma(\DsTTz)\mathcal{B}(\DsTTz\to X) \mathcal{B}( \Ds\to\phi\pip )$} \\
& $\Ds\pi^-$         & \multicolumn{3}{c}{---}     & $<  2.0$ \\
\hline
\multicolumn{6}{l}{$\sigma(\DsTS)\mathcal{B}(\DsTS\to X) \mathcal{B}( \Ds\to\phi\pip )$} \\
& $\Ds\pip\pim$      & $  5.2$ & $  0.7$ & $  0.4$ & ---\\ 
\end{tabular}
\end{ruledtabular}
{\footnotesize [a] Includes both $\DsTO\piz$ and $\DsTT\gamma$ channels.\hfill}
\end{table}

A 95\% CL upper limit is calculated for those yields which are not
statistically significant. These limits are calculated using a 
frequentist approach by determining in each case
the hypothetical value of $\bar{\sigma}$ that is 1.96 standard deviations
above the measured values. In order to calculate the
systematic uncertainty associated with
a hypothetical value of $\bar{\sigma}$, those uncertainties that
are proportional to the signal yield (such as uncertainties related
to selection efficiency) are scaled as appropriate. The results
are shown in Table~\ref{tb:yields}.

The yields listed in Table~\ref{tb:yields} are used to calculate
the branching ratios shown in Table~\ref{tb:branching}. 
For the $\DsTT$ meson, only one decay mode
has been observed; this is used as the denominator when calculating 
the $\DsTT$ branching ratios.
For the $\DsFE$ meson, the $\Ds\piz\gamma$ decay mode (consisting
of possible decay through either $\DsTO\piz$ or $\DsTT\gamma$) is chosen
for this role.
For completeness, the yield from hypothetical
$\DsTTdc\to\Ds\pip$ and $\DsTTz\to\Ds\pim$ decays is compared to
$\DsTT\to\Ds\piz$ decay to produce a quantity that is proportional
to both the respective branching ratios and production rates.

\begin{table}
\caption{\label{tb:branching}A summary of branching-ratio results.
The first quoted uncertainty for the central value
is statistical and the second is systematic. The limits
correspond to 95\% CL. For the hypothetical $\DsTTdc$ and $\DsTTz$ mesons,
an unknown additional factor from the ratio of production cross sections is
involved. A lower limit is quoted for the $\DsFE\to\DsTO\piz$ results.}
\begin{ruledtabular}
\renewcommand{\baselinestretch}{1.3}
\begin{tabular}{llr@{$\:\pm\:$}r@{$\:\pm\:$}rr}
\multicolumn{2}{l}{Decay Mode} 
     & \multicolumn{3}{c}{Central Value} & Limit  \\
\hline
\multicolumn{6}{l}{$\mathcal B(\DsTT\to X)/\mathcal B(\DsTT\to\Ds\piz)$} \\
\hspace{5pt}
& $\Ds\gamma$        & $-0.02$&$ 0.02$&$ 0.08$&$ <0.14$\\
& $\Ds\pi^0\pi^0$    & $ 0.08$&$ 0.06$&$ 0.04$&$ <0.25$\\
& $\Ds\gamma\gamma$  & $ 0.06$&$ 0.04$&$ 0.02$&$ <0.18$\\
& $\DsTO\gamma$      & $ 0.00$&$ 0.03$&$ 0.07$&$ <0.16$\\
& $\Ds\pi^+\pi^-$    & $ 0.0023$&$ 0.0013$&$ 0.0002$&$ <0.0050$\\
\hline
\multicolumn{6}{l}{$\mathcal B(\DsFE\to X)/\mathcal B(\DsFE\to\Ds\piz\gamma)$ [a]} \\
& $\Ds\piz$          & $-0.023$&$ 0.032$&$ 0.005$&$ <0.042$ \\
& $\Ds\gamma$        & $ 0.337$&$ 0.036$&$ 0.038$&    ---   \\
& $\DsTO\piz$        & $ 0.97$&$ 0.09$&$ 0.05$&$ >0.75$\\
& $\DsTT\gamma$      & $ 0.03$&$ 0.09$&$ 0.05$&$ <0.25$\\
& $\Ds\pi^0\pi^0$    & $ 0.13$&$ 0.13$&$ 0.06$&$ <0.68$\\
& $\Ds\gamma\gamma$  & $ 0.08$&$ 0.10$&$ 0.04$&$ <0.33$\\
& $\DsTO\gamma$      & $-0.02$&$ 0.08$&$ 0.10$&$ <0.24$\\
& $\Ds\pi^+\pi^-$    & $ 0.077$ & $0.013$  & $0.008$ & ---\\ 
\hline
\multicolumn{6}{l}{$\sigma(\DsTTdc)/\sigma(\DsTT) \times$} \\
\multicolumn{6}{l}{$\mathcal B(\DsTTdc\to X)/\mathcal B(\DsTT\to\Ds\piz)$} \\
& $\Ds\pi^+$         & \multicolumn{3}{c}{---} & $<0.017$ \\ 
\hline
\multicolumn{6}{l}{$\sigma(\DsTTz)/\sigma(\DsTT) \times$} \\
\multicolumn{6}{l}{$\mathcal B(\DsTTz\to X)/\mathcal B(\DsTT\to\Ds\piz)$} \\
& $\Ds\pi^-$         & \multicolumn{3}{c}{---} & $<0.013$ \\ 
\end{tabular}
\end{ruledtabular}
{\footnotesize [a] Denominator includes both 
$\DsTO\piz$ and $\DsTT\gamma$ channels.\hfill}
\end{table}

In order to calculate the systematic uncertainty in the branching ratio
results, systematic uncertainties common to the nominator and denominator
are first discarded. Such common uncertainties include those associated with
the $\Ds\to\phi\pip$ branching ratio correction and, in some cases,
those associated with $\piz$ and $\gamma$ reconstruction efficiency.

A 95\% CL upper limit is calculated for those branching ratios associated 
with decay modes that lack statistical signficance. The method used
is similar to the frequentist recipe used for the 
upper limits on the cross section. The results are included
in Table~\ref{tb:branching}.

\section{Discussion}

The results in this paper confirm previous measurements, with
generally higher precision. No new decay modes have been uncovered.
Lacking any additional evidence to the contrary, the $J^P=0^+$ and $J^P=1^+$
assignments for the $\DsTT$ and $\DsFE$ mesons remain a viable
hypothesis. Except, perhaps, for the mass, there is currently no 
conflict with the interpretation of the $\DsTT$ and $\DsFE$ mesons 
as the $J^P=0^+$ and $J^P=1^+$
$p$-wave $c\bar{s}$ states, as shown in Fig.~\ref{fg:godfreyisgur}.

The $\DsFE$ meson mass measured in this analysis is an
improvement over previous 
measurements~\cite{Besson:2003cp,Aubert:2003pe,Abe:2003jk}. 
High precision is obtained by using a decay mode that
includes only charged particles coupled with a detailed understanding
of the performance of the \babar\   tracking detectors. 
The determination of the $\DsTT$ mass remains limited by uncertainties
in EMC energy scale.

No intrinsic width of any statistical signficance is observed for either the
$\DsTT$ or $\DsFE$ meson. There is sufficient detector resolution
in the $\Ds\pip\pim$ decay mode that better limits for the $\DsFE$ meson
should be attainable in the future with higher statistics.
In contrast, limits on the $\DsTT$ width
are currently limited by systematic uncertainties and will be difficult to
improve.

Both the mass and width results for the $\DsTS$ meson 
obtained from the decay to $\Ds\pip\pim$
are consistent
with prior measurements~\cite{Eidelman:2004wy} and 
theoretical 
expectations~\cite{Godfrey:1985xj,Godfrey:1991wj,Isgur:1991wq,DiPierro:2001uu}.

The analysis of the decay $\DsFE\to\DsTO\piz$ is complicated
by the overlap in kinematics of the hypothetical decay 
$\DsFE\to\DsTT\gamma$. The analysis presented here applies a likelihood
technique to separate these two decay modes and provide a limit
on the latter.
A previous analysis from BELLE~\cite{Abe:2003jk}
of the decay $\DsFE\to\DsTO\piz$ did not account for 
such possible contamination.
As a consequence, it should be noted, their quoted yield for the
decay $\DsFE\to\DsTO\piz$ should be
compared to the $\DsFE\to\Ds\piz\gamma$
result presented here in which both decay channels ($\DsTT\gamma$
and $\DsTO\piz$) are included together. Specifically, the BELLE
measurement of the branching fraction from $e^+e^- \to c\overline{c}$
production is
$\mathcal B(\DsFE\to\Ds\gamma)/\mathcal B(\DsFE\to\DsTO\piz) 
= 0.55 \pm 0.13\;(\text{stat.}) \pm 0.08\;(\text{syst.})$
and can be compared directly to the result of this analysis
($0.337 \pm 0.036\;(\text{stat.}) \pm 0.038\;(\text{syst.})$).
The less precise BELLE result is somewhat larger, but still compatible
within uncertainties. The result presented here is consistent with
BELLE ($0.38 \pm 0.11\;(\text{stat.}) \pm 0.04\;(\text{syst.})$)~\cite{Krokovny:2003zq}
and
\babar\   ($0.275 \pm 0.045\;(\text{stat.}) \pm 0.020\;(\text{syst.})$)~\cite{Aubert:2004pw}
measurements from $\DsFE$ production in $B$ decay.

The remaining branching ratio results shown in Table~\ref{tb:branching} are
generally competitive with the corresponding BELLE limits~\cite{Abe:2003jk}.
An exception is the limit on the decay $\DsTT\to\Ds\gamma$.
The BELLE publication, however, does not address the difficulties
associated with modeling the $\DsTT\to\Ds\piz$ reflection (see
Fig.~\ref{fg:dsgam.plotfitdatapaper}) which is the source of much
of the systematic uncertainty quoted in this analysis. 

The searches
for the $\Ds\piz\piz$
and $\Ds\gamma\gamma$ decay modes of the $\DsTT$ and $\DsFE$ have been
reported in this analysis for the first time. Both final states suffer from
large backgrounds. More precise information on these modes will require
more advanced analysis techniques, perhaps with the use of $B$ meson
decay or some other method of background suppression.

The $\DsTT$ and $\DsFE$ masses are considerably lower 
than predictions from
potential models developed before their discovery.
Since the $\DsTT$/$\DsFE$ mass splitting is so much larger than that
for the $\DsTS$ and $\DsST$ mesons, it would appear that this conflict 
is intrinsic. Studies from Cahn and Jackson~\cite{Cahn:2003cw} suggest,
however, that it is possible with some adjustment to fit the
$\DsTT$ and $\DsFE$ into a perturbative model. In any case, given
the approximate nature of these models, there is a danger
that even large discrepancies can be overstated.

In contrast, according to Bardeen, Eichten, and Hill,
models based on chiral symmetry naturally accommodate
the large $\DsTT$/$\DsFE$ mass splitting~\cite{Bardeen:2003kt} when
the $\DsTT$ and $\DsFE$ multiplet is treated as a chiral partner
of the ground state $\Ds$ and $\DsTO$ mesons. 
Their model provides specific predictions for branching
ratios, notably 
$\mathcal B(\DsFE\to\Ds\pip\pim)/\mathcal B(\DsFE\to\DsTO\piz) \approx 0.09$
and
$\mathcal B(\DsFE\to\Ds\gamma)/\mathcal B(\DsFE\to\DsTO\piz) \approx 0.24$,
both in good agreement with our measurements of
($0.077 \pm 0.015$) and ($0.337 \pm 0.052$), respectively.
Other branching ratio predictions are
$\mathcal B(\DsTT\to\DsTO\gamma)/\mathcal B(\DsTT\to\Ds\piz) \approx 0.08$
and
$\mathcal B(\DsFE\to\DsTT\gamma)/\mathcal B(\DsFE\to\DsTO\piz) \approx 0.13$.
Neither decay is observed by this analysis, but the corresponding
upper limits still exceed the predicted branching ratios.
This model also predicts a total intrinsic width of 23.2~\kev\  and
38.2~\kev\  for the $\DsTT$ and $\DsFE$ mesons, respectively,
far below the limits measured here.

Perturbative calculations from Godfrey~\cite{Godfrey:2003kg} 
predict branching ratios of $0.19$ and $0.55$ for
the radiative transition decay of the $\DsTT$ and $\DsFE$ to $\DsTO\gamma$.
These predictions are not consistent with the results reported here.
This discrepancy could be considered evidence that both the $\DsTT$
and $\DsFE$ are not $c\bar{s}$ mesons. These theoretical predictions,
however, are plagued with large
uncertainties in the partial widths of the isospin-violating
$\DsTT\to\Ds\piz$ and $\DsFE\to\DsTO\piz$ decays.
In addition, leading-order corrections to the radiative transition
partial rates might explain current experimental limits~\cite{Mehen:2004uj}.
Nevertheless, the possibility that the $\DsTT$ and $\DsFE$ are
some type of new bound state, such as a four-quark $DK$ molecule,
should be taken seriously.

The possibility of four-quark states has long been 
proposed~\cite{Jaffe:1976ih,Lipkin:1977ie}. Perhaps the most
unambiguous signature of the molecular interpretation of the $\DsTT$ meson
would be the production of neutral and
charged partners decaying to $\Ds\pipm$. For an isospin 1 $DK$
molecule, production of these isospin partners is expected
at the same rate as the $\DsTT$~\cite{Barnes:2003dj}.
This is clearly ruled out by our data. Other molecular interpretations,
however, do not have this limitation~\cite{Barnes:2003dj,Lipkin:2003zk},
and it may be challenging to rule them all out with the data 
from the $B$-factories alone.

Finally, now that more precise data on the $\DsTT$ and $\DsFE$ is available,
measurements of the other $c\bar{s}$ states have become more
important. The measurement of the $\DsTS$ mass in this analysis
will be helpful in further constraining models. With high statistics
samples available from both $B$-factory experiments, improvements
in the parameters of the $\Ds$, $\DsTO$, and $\DsST$ mesons should not
be neglected and will hopefully follow soon.

\section{Conclusion}

An updated analysis of the $\DsTT$ and $\DsFE$ mesons using
232~${\rm fb}^{-1}$ of $e^+e^-\to c\bar{c}$ data is performed.
Established signals from the decay $\DsTT\to\Ds\piz$ and 
$\DsFE\to\Ds\piz\gamma$, $\Ds\gamma$, and $\Ds\pip\pim$ are confirmed.
A detailed analysis of invariant mass distributions of these
final states including consideration of the background introduced
by reflections of other $c\bar{s}$ decays
produces the following mass values:
\begin{equation*}
\renewcommand{\arraystretch}{1.2}
\begin{array}{lcl@{}r@{\:}c@{\:}r@{\:}c@{\:}r@{}l}
m(\DsTT)
&=& (&2319.6 &\pm&   0.2  &\pm&   1.4&)
\;\mevcc  \\
m(\DsFE)
&=& (&2460.1 &\pm&   0.2  &\pm&   0.8&)
\;\mevcc,
\end{array}
\end{equation*}
where the first error is statistical and the second systematic.
Upper 95\% CL limits of $\Gamma < 3.8$~\mev\  and $\Gamma < 3.5$~\mev\  
are calculated for the intrinsic $\DsTT$ and $\DsFE$ widths.
All results are consistent with previous measurements.

The following final states are investigated: $\Ds\piz$, $\Ds\gamma$,
$\DsTO\piz$, $\DsTT\gamma$, $\Ds\piz\piz$, $\DsTO\gamma$,
$\Ds\gamma\gamma$, $\Ds\pipm$, and $\Ds\pip\pim$. No 
statistically significant evidence of new decay modes is observed.
The following branching ratios are measured:
\begin{equation*}
\renewcommand{\arraystretch}{1.8}
\begin{array}{ccl@{}r@{\:}c@{\:}r@{\:}c@{\:}r}
\frac{\mathcal B(\DsFE\to\Ds\gamma)}{\mathcal B(\DsFE\to\Ds\piz\gamma)}
&=& &0.337 &\pm&   0.036  &\pm&   0.038 \\
\frac{\mathcal B(\DsFE\to\Ds\pip\pim)}{\mathcal B(\DsFE\to\Ds\piz\gamma)}
&=& &0.077 &\pm&   0.013  &\pm&    0.008
\end{array}
\end{equation*}
where the first error is statistical and the second systematic.
The data are consistent with the decay $\DsFE\to\Ds\piz\gamma$ 
proceeding entirely through $\DsTO\piz$.

Since the results presented here are consistent with $J^P=0^+$
and $J^P = 1^+$ spin-parity assignments for the $\DsTT$ and
$\DsFE$ mesons, these two states remain viable candidates
for the lowest lying $p$-wave $c\bar{s}$ mesons.
The lack of evidence for some radiative decays,
in particular $\DsTT\to\DsTO\gamma$ and $\DsFE\to\DsTO\gamma$,
are in contradiction with this hypothesis according to some
calculations, but large theoretical uncertainties remain.
No state near the $\DsTT$ mass is observed decaying to $\Ds\pipm$.
If charged or neutral partners to the $\DsTT$ exist (as would be
expected if the $\DsTT$ is a four-quark state), some mechanism
is required to suppress their production in $e^+e^-$ collisions.

~  

\begin{acknowledgments}
We are grateful for the 
extraordinary contributions of our \pep2\ colleagues in
achieving the excellent luminosity and machine conditions
that have made this work possible.
The success of this project also relies critically on the 
expertise and dedication of the computing organizations that 
support \babar.
The collaborating institutions wish to thank 
SLAC for its support and the kind hospitality extended to them. 
This work is supported by the
US Department of Energy
and National Science Foundation, the
Natural Sciences and Engineering Research Council (Canada),
Institute of High Energy Physics (China), the
Commissariat \`a l'Energie Atomique and
Institut National de Physique Nucl\'eaire et de Physique des Particules
(France), the
Bundesministerium f\"ur Bildung und Forschung and
Deutsche Forschungsgemeinschaft
(Germany), the
Istituto Nazionale di Fisica Nucleare (Italy),
the Foundation for Fundamental Research on Matter (The Netherlands),
the Research Council of Norway, the
Ministry of Science and Technology of the Russian Federation, and the
Particle Physics and Astronomy Research Council (United Kingdom). 
Individuals have received support from 
CONACyT (Mexico), the Marie-Curie Intra European Fellowship program (European Union),
the A. P. Sloan Foundation, 
the Research Corporation,
and the Alexander von Humboldt Foundation.

\end{acknowledgments}

\bibliography{note1167}

\begin{thebibliography}{22}
\bibitem[{cpf()}]{cpfootnote}
\bibinfo{note}{Inclusion of charge conjugate states is implied throughout this
  paper.}
\end{thebibliography}

\end{document}